\documentclass{aa}  

\usepackage{graphicx}
\usepackage{txfonts}
\usepackage{dsfont}
\usepackage{xcolor}
\usepackage{subcaption}
\usepackage{booktabs}
\usepackage{array}
\usepackage{amsmath}
\usepackage{placeins}
\newcolumntype{E}{>{$\,\displaystyle}l<{$}} 

\usepackage[colorlinks = true,citecolor = blue, linkcolor = blue, urlcolor = blue]{hyperref}

\begin{document}

   \title{YOLO-CIANNA: Galaxy detection with deep learning in radio data}

   \subtitle{II. Winning the SKA SDC2 using a generalized 3D-YOLO network}
   
   \titlerunning{YOLO-CIANNA: Galaxy detection with deep learning in radio data}
    
   \author{D. Cornu
          \inst{1}, B. Semelin \inst{1}, P. Salomé \inst{1}, X. Lu \inst{6}, S. Aicardi \inst{5}, J. Freundlich \inst{4}, \break
           F. Mertens \inst{1}, A. Marchal \inst{2,3}, G. Sainton \inst{1}, F. Combes \inst{1,7}, C. Tasse \inst{1,9}}
    \authorrunning{Cornu et al.}

   \institute{$^{1}$LUX, Observatoire de Paris, Université PSL, Sorbonne Université, CNRS, 75014, Paris, France\\
$^{2}$Canadian Institute for Theoretical Astrophysics, University of Toronto, 60 St. George Street, Toronto, ON M5S 3H8 \\
$^{3}$Research School of Astronomy \& Astrophysics, Australian National 
University, Canberra ACT 2610 Australia \\
$^{4}$Université de Strasbourg, CNRS UMR 7550, Observatoire astronomique de Strasbourg, 67000 Strasbourg, France\\
$^{5}$DIO, Observatoire de Paris, CNRS, PSL, 75014, Paris, France\\
$^{6}$IDRIS, CNRS, F-91403 Orsay, France\\
$^{7}$Collège de France, 11 Place Marcelin Berthelot, 75005, Paris, France\\
$^{9}$Department of Physics \& Electronics, Rhodes University, PO Box 94, Grahamstown, 6140, South Africa\\
}
\date{Received September 15, 2025; accepted January 26, 2026}

  \abstract
    {As the scientific exploitation of the Square Kilometre Array (SKA) approaches, there is a need for new advanced data analysis and visualization tools capable of processing large high-dimensional datasets.}
   {In this study, we aim to generalize the YOLO-CIANNA deep learning source detection and characterization method for 3D hyperspectral HI emission cubes.}
   {We present the adaptations we made to the regression-based detection formalism and the construction of an end-to-end 3D convolutional neural network (CNN) backbone. We then describe a processing pipeline for applying the method to simulated 3D HI cubes from the SKA Observatory Science Data Challenge 2 (SDC2) dataset.}
    {The YOLO-CIANNA method was originally developed and used by the MINERVA team that won the official SDC2 competition. Despite the public release of the full SDC2 dataset, no published result has yet surpassed MINERVA's top score. In this paper, we present an updated version of our method that improves our challenge score by 9.5\%. The resulting catalog exhibits a high detection purity of 92.3\%, best-in-class characterization accuracy, and contains 45\% more confirmed sources than concurrent classical detection tools. The method is also computationally efficient, processing the full $\sim$1TB SDC2 data cube in 30 min on a single GPU.}
   {These state-of-the-art results highlight the effectiveness of 3D CNN-based detectors for processing large hyperspectral data cubes and represent a promising step toward applying YOLO-CIANNA to observational data from SKA and its precursors.}

   \keywords{
            Methods: numerical --
            Methods: statistical --
            Methods: data analysis --
            Galaxies: statistics --
            Radio lines: galaxies
            }

   \maketitle
%
\section{Introduction}
\label{sec:introduction}

New observing facilities generate ever larger datasets, which are often high-dimensional and have a wide dynamic range. The combined volume and complexity of these datasets put stress on classical data processing pipelines, which often fail to scale to the required data rate. To overcome these difficulties, statistical approaches such as machine learning (ML) methods are increasingly employed, as they are known to scale well with data size and dimensionality. The Square Kilometer Array \citep[SKA,][]{paper:ska_ref} is a prominent example of next-generation telescopes in the radio domain. This instrument is expected to transform our understanding of the Universe, with a wide range of applications from setting constraints on the cosmic dawn to the detection of new exoplanets in our Galaxy. The unprecedented data rate foreseen for SKA is already prompting the international community to develop new strategies for data processing, storage, and distribution \citep{paper:scaife_ska}. In this context, the SKA Observatory (SKAO) organizes recurrent Science Data Challenges (SDCs) over simulated SKA-like data products \citep{paper:sdc1, paper:sdc2, paper:sdc3a}. These challenges aim to stimulate the development of highly efficient methods and to enable the community to prepare for handling SKA data.

The present paper is the second in a series introducing YOLO-CIANNA, a deep-learning regression-based method for source detection and characterization. The series aims to evaluate the method's performance on simulated data from the SDCs, in preparation for its applications to observational data from SKA precursors and pathfinders. The first paper, \citet[][]{paper:yolo_cianna_sdc1} hereafter Paper I, covers the full method description and its application to simulated continuum images from the SDC1 dataset \citep{paper:sdc1}. It also includes a concise review of ML object detection methods along with some examples of astronomical applications. The method description emphasizes specific design choices tailored to astronomical datasets. This first paper concludes with a discussion of the biases and limitations associated with the SDC1 challenge definition, and presents possible improvements to the method itself. 

In the present work, we extend the results of Paper I by generalizing YOLO-CIANNA to 3D source detection in hyperspectral data and apply it to simulated HI emission cubes from the SDC2. The method was initially developed and used by team MINERVA (Machine Learning for Radioastronomy at Observatoire de Paris) for its participation in SDC2, in which the team secured first place. Therefore, this paper provides a detailed description of an updated version of MINERVA's method.

\section{SDC2 description}
\label{sec:data_descrption}

\subsection{Data cubes and source catalogs}

The SDC2 is a galaxy detection and characterization challenge in simulated 3D HI cubes representative of future SKA-mid observations. A complete description of the challenge and of the associated data can be found in the SDC2 summary paper \citet{paper:sdc2}, which details the simulation of the source catalog using \texttt{TRECS} \citep{paper:t-recs}, the construction of the sky model, and the addition of instrumental effects. The primary data product is a MAIN cube covering a 20 square degree region with a synthesized beam of 7 arcsec and a pixel resolution of 2.8 arcsec. It covers a frequency bandwidth of 950-1150 MHz ($z=0.235$-0.495) sampled at 30 kHz. This results in a massive cube (851 GB) that is $5851 \times 5851 \times 6668$ (pixels, pixels, channels) in size, corresponding to $228 \times 10^9$ voxels, a voxel being defined as the volume of a square pixel over one frequency channel. Continuum images of the same field are also provided with identical spatial resolution and bandwidth, but sampled at a 10 MHz resolution. Thermal noise consistent with 2000h of SKA integration time was added, yielding 26 to 31 {\textmu Jy}~$\mbox{beam}^{-1}$ per channel. The cube also features continuum subtraction residues, instrumental artifacts, and radio frequency interference (RFI) subtraction residues, improving its realism.

The MAIN cube contains 233\,245 simulated HI sources. Each is characterized by its sky coordinates (RA, Dec), integrated line flux, $f$, HI major axis diameter, $S$, at 1 \mbox{M$_\odot\,$pc$^{-2}$} (hereafter HI size), line width, $w_{20}$, measured at 20\% of the peak, major axis position angle, $PA$, and inclination angle, $i$ (between the line of sight and the galaxy plane). These parameters are known for each source and stored in a truth catalog that was kept secret during the original challenge but released afterward. Instead, participating teams were provided with a smaller LDEV cube (40 GB), obtained through the same data generation procedure, but with a reduced field of view of 1 square degree, and for which the truth catalog was provided. This LDEV cube has the same spatial and spectral resolution as the MAIN cube and covers the same bandwidth, resulting in a size of $1286 \times 1286 \times 6668$. It is also accompanied by continuum images of the corresponding field. The associated LDEV truth catalog contains 11\,091 sources and can be used to calibrate or train detection methods. In this paper, we reproduced the challenge conditions by training our detector on the LDEV cube and catalog. The now-released MAIN truth catalog is used to analyze our results and compare them with the SDC2 summary paper.

\begin{figure}
    \centering
    \includegraphics[width=1.0\hsize]{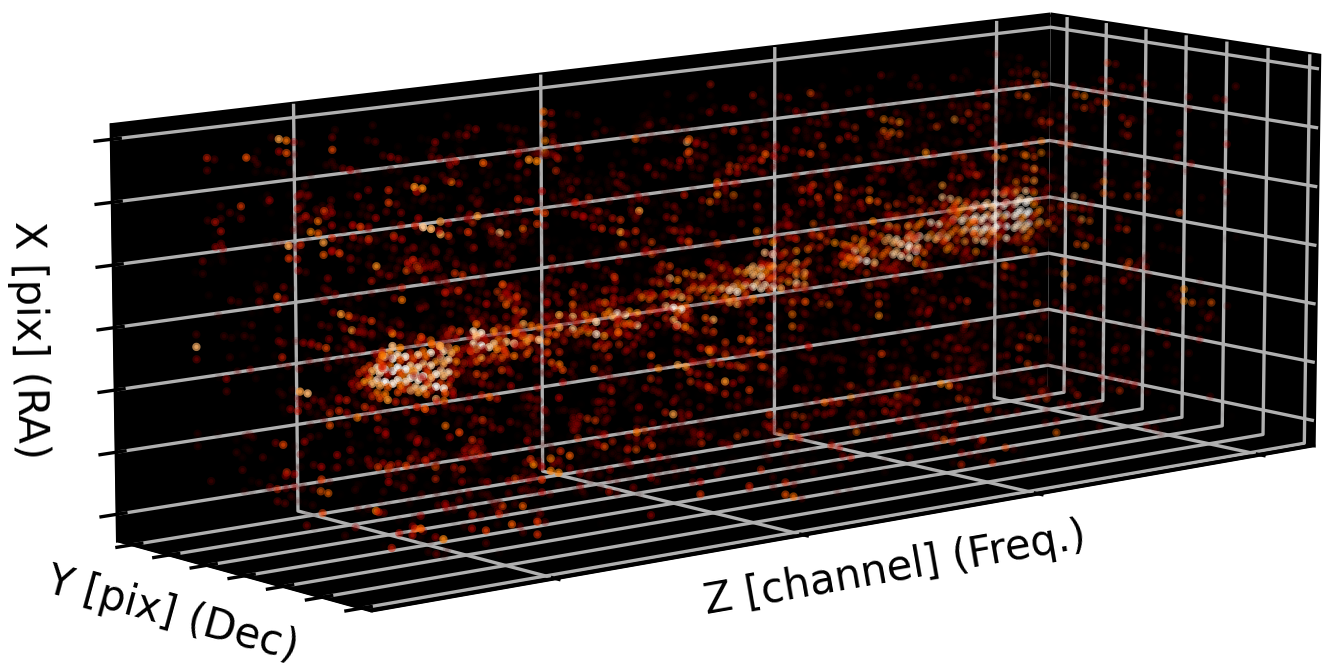}
    \caption{3D view of a $30\times 30\times 70$ subcube centered on a bright source from the MAIN catalog (id. 15\,770) using a rescaled intensity. Transparency is inversely proportional to voxel intensity.}
    \label{fig:bright_source_3d}
\end{figure}

Figure~\ref{fig:bright_source_3d} shows a cutout around one of the brightest sources from the truth catalog. To improve contrast, voxel values have been rescaled following \mbox{$p'_i = \tanh\left( 3 p_i / \max_p \right)$}, where $p_i$ and $p'_i$ are the original and rescaled voxel values, respectively, and $max_p$ is the maximum voxel value in the cutout. To better visualize the inner 3D structure, we set voxel transparency to be inversely proportional to their rescaled value. This figure illustrates the typical 3D structure of a galaxy observed in HI, where the line width correlates with its rotation. With the SDC2 observational setup, sources are compact in the plane of the sky but extended over tens of channels, with lower signal intensity at the central coordinates. The signal is mostly aligned in a plane, which defines the source angles, $PA$ and $i$.

\subsection{SDC2 scoring metric}

The challenge task is to produce a catalog of detected sources in the MAIN cube along with all their properties. To evaluate the quality of a submission, the SDC2 relies on a dedicated scoring system. We provide a summary of the score computation as implemented in the scorer code\footnote{\href{https://gitlab.com/ska-telescope/sdc/ska-sdc}{https://gitlab.com/ska-telescope/sdc/ska-sdc}}. The SDC2 scorer adopts the SDC1 matching system, which adds size and flux accuracy criteria to the classical sky separation metric. It is generalized to hyperspectral data by adding line width and central frequency accuracy criteria. The scorer defines a multiparameter error as
\begin{align}
D_{\rm tot} &= \sqrt{D^2_{\textrm{pos}} + D^2_{\textrm{freq}} + D^2_{\textrm{flux}} + D^2_{\textrm{size}} + D^2_{\textrm{width}}} \text{, with} \label{eq:match_tot}\\
D_{\textrm{pos}} &= \sqrt{(x-\hat{x})^2 + (y-\hat{y})^2}/\hat{S}'\text{,}\\
D_{\textrm{freq}} &= \left|\nu-\hat{\nu}\right|/\hat{w}_{20,Hz}\text{,}\\
D_{\textrm{flux}} &= \left|f - \hat{f}\right|/\hat{f},\\
D_{\textrm{size}} &= \left|S - \hat{S}\right|/\hat{S}'\text{, and}\\
D_{\textrm{width}} &= \left|w_{20} - \hat{w}_{20}\right|/\hat{w}_{20},\label{eq:match_width}
\end{align}
where truth values are indicated with a hat, ($x$,$y$) are the central pixel coordinates (converted from RA, Dec), $\nu$ the central frequency, $S'$ the HI size convolved with the synthesized beam, $f$ the line flux integral, $S$ the HI size, and $w_{20}$ the line width. Unlike the SDC1, all terms are equally weighted, and the global rejection threshold is set to $D_{\rm tot} \geq 5$. However, there are additional conditions on both the position and frequency that reject sources if $D_{\textrm{pos}} \geq 1$ or $D_{\textrm{freq}} \geq 1$. If multiple predictions match the same truth source, all are retained in the match catalog. However, their contribution to the final score is weighted down by the number of matches. In the rare case of a single prediction matching multiple truth sources, it is associated with the lowest $D_{\rm tot}$.

Matches are scored individually by comparing the predicted and true characteristics. The scorer defines a subscore for each source property to predict. Their average gives the match score. These are expressed as
\begin{equation}
s_k^j = \min{\left(1,\frac{T^j}{E_k^j}\right)}, \quad {\rm and} \quad s_k = \frac{1}{7} \sum_j^7 s_k^j,
\label{eq:score_response_fct}
\end{equation}
where $j$ represents a subscore, $k$ identifies one source in the detection catalog, $T^j$ is a threshold specific to each subscore, $E_k^j$ is the error term specific to each subscore (possibly different from the $D$ terms used as matching criteria), and $s^j$ is the resulting subscore. Error functions for each parameter are listed in Table~\ref{table:error_functions}. The final individual score is bound to the $[0,1]$ interval. The total score is the sum of all individual matched scores minus the number of false detections, following
\begin{equation}
M_s = \sum_k^{N_{\textrm{match}}}{s_k} - N_{\rm false}. \label{eq:total_score}
\end{equation}
In addition, we can also estimate mean subscores and a global average characterization score as
\begin{equation}
\bar{s}^j = \frac{1}{N_{\textrm{match}}}\sum_k^{N_{\textrm{match}}} s_k^j \quad \text{ and } \quad \bar{s} = \frac{1}{N_{\textrm{match}}}\sum_k^{N_{\textrm{match}}}{s_k}.
\end{equation}
The scorer can be used for either the MAIN or LDEV cube. Hereafter, the term score refers to a scoring over the MAIN cube, unless stated otherwise.

\begin{table}
\centering
\caption{\label{table:error_functions} Error functions and thresholds for all subscores.}
\begin{tabular}{ l E @{=} E c }
 \hline\hline
 Subscore & \multicolumn{2}{c}{Error function $E^j$} & $T^j$ \\
 \hline\\[-2.5ex]
 Sky position      & E^{\textrm{pos}}   & \,\sqrt{(x-\hat{x})^2 +(y-\hat{y})^2}/\hat{S}'' & 0.3 \\
 Central frequency & E^{\textrm{freq}}  & \,| \nu - \hat{\nu} |/\hat{w}_{20,Hz} & 0.3 \\
 Line flux         & E^{\textrm{flux}}  & \,| f - \hat{f} |/\hat{f} & 0.1 \\
 HI size           & E^{\textrm{size}}  & \,| S - \hat{S} |/\hat{S}'' & 0.3 \\
 Line width        & E^{\textrm{width}} & \,| w_{20} - \hat{w}_{20}|/\hat{w}_{20} & 0.3 \\
 Position angle    & E^{PA}             & \,| PA - \hat{PA}| & 10.0 \\
 Inclination angle & E^{\textrm{inc}}   & \,| i - \hat{i}| & 10.0 \\
 \hline
\end{tabular}\\
{\raggedright \vspace{0.2cm} Note: $\hat{S}''$ is the target HI size convolved with twice the synthesized beam size, which is different from $\hat{S}'$.\par}
\end{table}

\subsection{Detectability selection function}
\label{sec:selection_function}

YOLO-CIANNA is a supervised ML approach and therefore requires a set of labeled examples. For this, we relied on the provided LDEV cube and truth source catalog. However, as with the SDC1, the vast majority of the listed sources are too faint to be detectable ($\sim85\%$). Including undetectable sources in the target catalog may be debatable from a challenge perspective. Still, the presence of faint sources in the simulation enhances background noise realism. It also ensures a realistic source signal-to-noise ratio (S/N) distribution in the target catalog without imposing an arbitrary detectability cut.

\begin{figure}
    \centering
    \includegraphics[width=1.0\hsize]{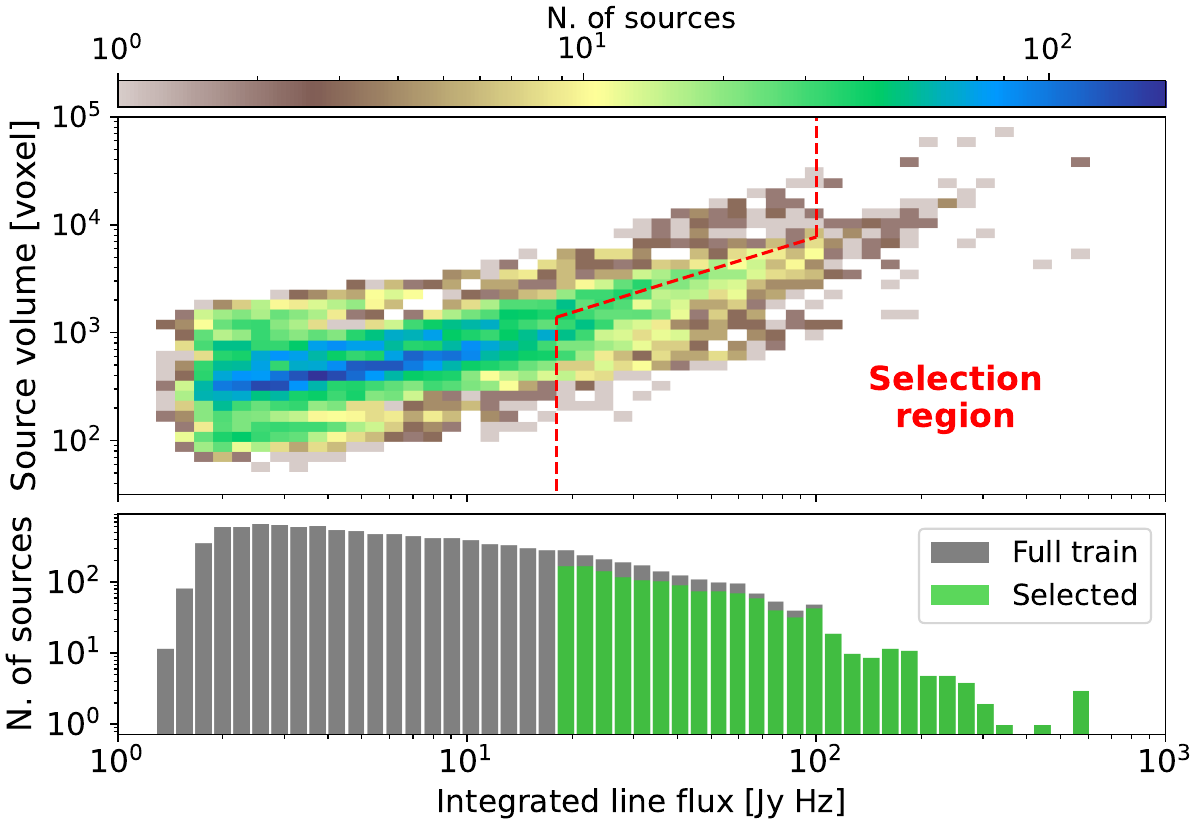}
    \caption{\textit{Top:} Two-dimensional histogram of LDEV truth source volume as a function of their integrated line flux. The dashed red line represents our selection function, with sources to the right of the line being selected. \textit{Bottom:} Line flux distribution for the full LDEV catalog and the selected sample.}
    \label{fig:selection_function}
\end{figure}

As with most supervised ML approaches, our method is sensitive to mislabeling in the training examples. It involves objects that should be detectable but not labeled as targets, or labeled targets that are too faint to be distinguishable from the background (Sects.~2.2 and 3.4 of Paper I). Therefore, we defined a detectability selection function to curate the LDEV truth catalog. This function combines the integrated line flux, $f$, with an estimate of the source volume brightness, $V_b$. We approximate the source volume by a cylinder with a radius equal to its beam-convolved HI size, expressed in pixels, and a length equal to its line width, in channels. The volume brightness was obtained as
\begin{align}
V_b &= \frac{\hat{f}}{\pi S'^2D} \text{, with} \label{eq:source_volume_brightness}\\
S' &= \sqrt{\hat{S}_{deg}^2 + B_s^2}\frac{1}{P_s}, \quad \textrm{ and} \label{eq:source_size} \quad
D = \frac{\hat{w}_{20}}{c}\frac{\hat{\nu}^2}{\nu_{HI_0}}\frac{1}{C_s},
\end{align}
where $\hat{S}_{deg}$ is the target source size in degree, $B_s$ the synthesized beam size in degree, $P_s$ the pixel size in degree, $\hat{w}_{20}$ the line width and $c$ the speed of light in kilometer per second, $\hat{\nu}$ the target central frequency in Hertz, $\nu_{HI_0}$ the rest frequency of HI in Hertz, and $C_s$ the frequency channel size in Hertz. The resulting volume brightness is in Jansky Hertz voxel$^{-1}$. The selection function is then defined as
\begin{equation}
    \big(f \geq 100\big) \quad \textrm{or} \quad \big(f > 18 \; \text{and}\; V_b > 0.013 \big).
\label{eq:selection_function}
\end{equation}
The exact flux and volume brightness thresholds are the result of an extensive manual search aimed at maximizing the score of a detector trained on the corresponding training sample. When applied to the LDEV truth catalog, this function flags 1383 of the 11\,091 sources as detectable and thus usable as training targets. We represent our selection cuts over a 2D histogram of the source volume against the line flux integral for the LDEV truth catalog in Fig.~\ref{fig:selection_function}, along with a comparison of the line flux histograms for the truth catalog with our selected training catalog.

\subsection{Data cube preprocessing}
\label{sec:cube_normalization}

We performed a series of preprocessing steps to enhance source detectability and prepared the data cube for use with our detector. We started by identifying the sky positions for which the mean signal over all continuum images is above $3\times10^{-3}$ $\mbox{Jy}\,\mbox{beam}^{-1}$ and set all the corresponding cube voxels to zero, masking continuum-subtraction residuals. We then normalized every frequency channel by its standard deviation. This reduces the impact of RFI subtraction residuals and makes the channel noise homogeneous across the whole bandwidth. It also allows us to consider the detection task as invariant with respect to the frequency position. This way, a detector can be trained and applied over subcubes along the frequency axis.

Due to the high frequency resolution, the source signal is diluted over multiple channels. This could be mitigated by smoothing the cube along the frequency axis or averaging every few channels, which would increase the apparent S/N for some sources. However, it would destroy part of the fine-grained 3D structure that the detector could exploit. We observed that such preprocessing allows for faster training convergence but reduces the final score. We therefore retained the original frequency resolution to maximize detection performance.

The last step was to renormalize the cube. As for the SDC1, we applied a scaled hyperbolic tangent transformation to the voxels following $p_i' = \tanh{\left( \alpha p_i\right)}$, where $p_i$ is the voxel value after the previous preprocessing steps, and $p_i'$ is the final normalized voxel value. The parameter $\alpha$ controls the steepness of the curve and thus impacts the final signal contrast and saturation. After some exploration, we adopted $\alpha=0.3$ for the SDC2. We normalized both the MAIN and LDEV cubes once and stored them in a reduced unsigned 16-bit integer format for faster dynamic loading. We observed no impact of this reduction on the final score, provided the reduction is done after the tanh normalization. To further reduce the loading time, we split the MAIN cube into chunks of $512 \times 512 \times 6668$ (3.5 GB), cutting along the two sky axes. The cuts were made based on our prediction pipeline setup (Sect.~\ref{sec:prediction_pipeline}), ensuring partial overlap and padding when necessary. The LDEV cube was left as a single file and fully loaded into system memory for training (Sect.~\ref{sec:training_example_gen}).

\section{Method}
\label{sec:method}

\subsection{YOLO-CIANNA generalization for 3D detection}
\label{sec:yolo_cianna_desc}

In Paper I \citep{paper:yolo_cianna_sdc1}, we provide a complete description of the YOLO-CIANNA method of detecting, classifying, and characterizing objects in 2D images. Here, we present adaptations of the formalism for detecting 3D objects. As is explained in Paper I, YOLO-CIANNA is a regression-based detection method implemented as part of the broader \texttt{CIANNA} development framework \citep{soft:cianna}\footnote{CIANNA is open-source and freely accessible through GitHub \href{https://github.com/Deyht/CIANNA}{https://github.com/Deyht/CIANNA}. The version used in this paper corresponds to the V-1.0 release \href{https://doi.org/10.5281/zenodo.12806325}{10.5281/zenodo.12806325}.}. The YOLO-CIANNA method constrains a fully convolutional neural network (CNN) backbone to create a mapping from an input data space to a regular grid of detection units. Each unit is associated with an independent subregion of the input space and is tasked with detecting all objects whose central coordinate lies within that region. Objects are represented as bounding boxes. During the training phase, the network is fed with labeled images. Each detection unit produces a set of candidate boxes, which are compared to the known list of target boxes through a matching function. The network weights are then adjusted to minimize the observed difference between its prediction and the expected result. Additionally, the model learns to predict an objectness score for each detection, representing its confidence that a given predicted box corresponds to a real object.

To generalize the formalism presented in Paper I to 3D detection, we added two new dimensions to the detection units' output vector to predict both the box's central position, $z$, and its size, $d$, along the third dimension. We express a 3D bounding box as a function of output vector elements, following
\begin{align}
        x = o^x + g_x\,, \quad\quad y = o^y + g_y\,, \quad\quad z = o^z + g_z, \label{eq:pos_output}\\
        w = p_w \mathrm{e}^{(o^w)}\,, \quad\quad h = p_h \mathrm{e}^{(o^h)}\,, \quad\quad d = p_d \mathrm{e}^{(o^d)},\label{eq:size_output}
\end{align}
where $g_x$, $g_y$, and $g_z$ are the output grid coordinates of the unit making the prediction, $o^x$, $o^y$, and $o^z$ are predicted sigmoid-activated values refining the position inside the subregion, $p_w$, $p_h$, and $p_d$ are predefined size priors, and $o^w$, $o^h$ and $o^d$ are preditect linear-activated values adjusting the size of the predicted box. The corresponding $\mathcal{L}_{\textrm{pos}}$ and $\mathcal{L}_{\textrm{size}}$ regression sublosses from Eq.~14 of Paper I are trivially adapted with a third dimension to obtain our full YOLO-CIANNA 3D loss.

We also updated the box proximity and similarity metric used in many parts of the method. For this, we relied on the intersection over union (IoU) measurement generalized to 3D boxes. In this paper, we use its distance-aware variant, the ${\rm DIoU}$ \citep{paper:diou}, bounded within $[-1,1]$, and defined as
\begin{equation}
    {\rm DIoU} = {\rm IoU} - \frac{\rho^2}{d^2},
\label{eq:diou}
\end{equation}
where $\rho$ is the Euclidean distance between the two box centers and $d$ is the diagonal length (in 3D, the space diagonal) of the smallest box that encloses the two boxes to compare. This proximity criterion is used both to define the target objectness and in the association function, as described in Paper I.

\subsection{Network backbone architecture}
\label{sec:network_backbone}

Using the updated method, a classical 2D CNN backbone could be used for 3D detection by treating the third dimension as a stack of independent 2D input channels. However, taking advantage of spatial invariance through fine-grained gridding across all positional dimensions usually improves detection accuracy. Thus, we gridded the third dimension into independent detection units as well. For this, we relied on \texttt{CIANNA}'s ability to build 3D spatial convolutional layers and design a fully 3D network backbone. The output layer is therefore directly in the form of a 3D output grid (indexed by $g_x$, $g_y$, $g_z$), and detection units are structurally centered on their associated input subvolume. This structure also reduces the number of learnable weights required in the model compared to a 2D backbone approach. More details about the interplay between our method and the network backbone can be found in Sect.~3.6 and Appendix~A of Paper I.

\begin{figure}
    \centering
    \includegraphics[width=1.0\hsize]{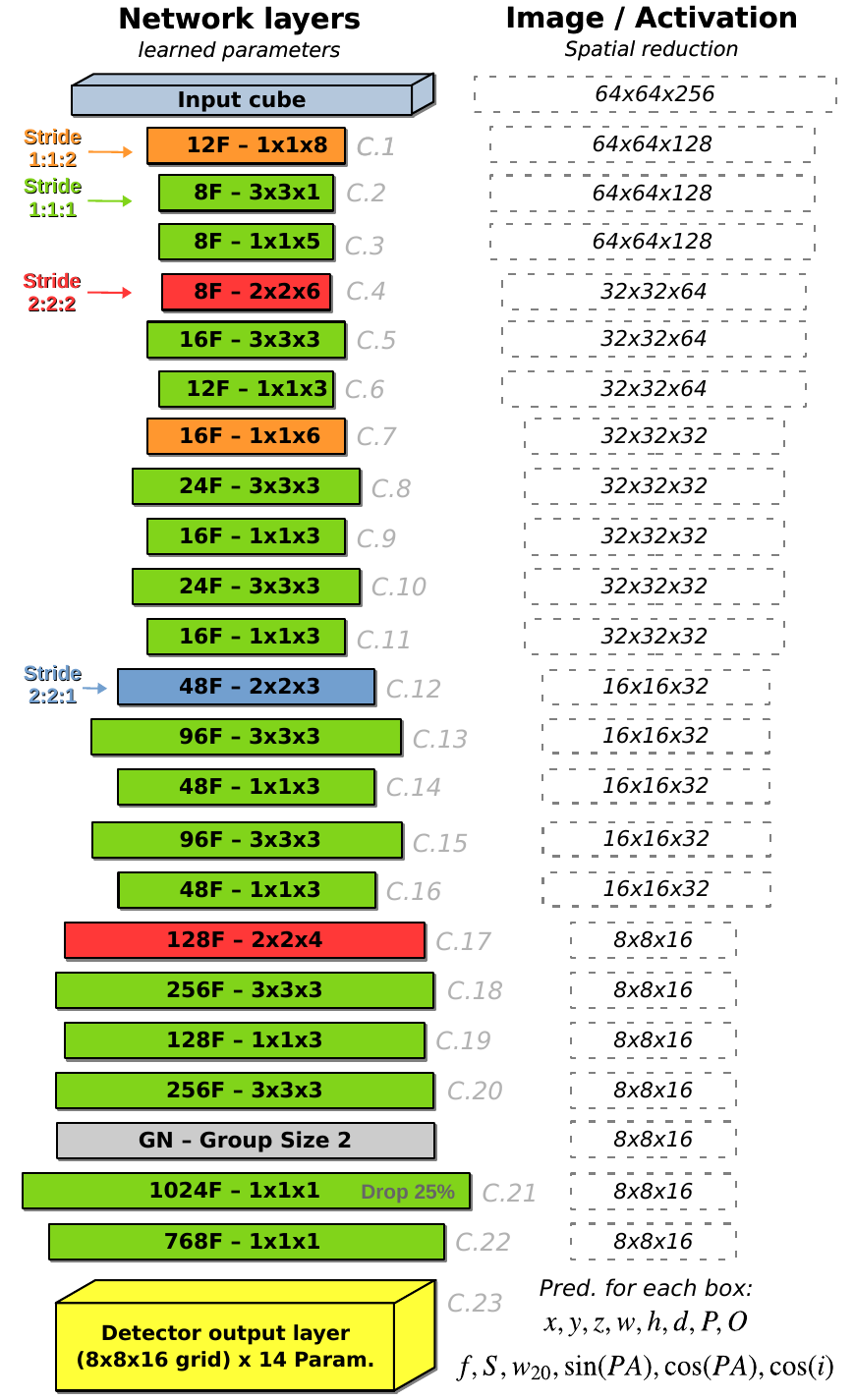}
    \caption{Our SDC2 3D CNN backbone architecture. The \textit{left} column provides layer structural properties, whereas the \textit{right} column indicates the spatial output dimension starting from a $64 \times 64 \times 256$ input cube. Layers are stacked top to bottom. The layer width scales with the number of filters. Layer colors encode the stride setting: green preserves dimensions, orange reduces the frequency dimension by two, red reduces all dimensions by two, and blue reduces sky dimensions by two. The output grid size and the predicted parameters are shown next to the last layer.}
    \label{fig:network_architecture}
\end{figure}

Even though a fully convolutional architecture is not bound to a specific input size, we chose to use a fixed subcube input dimension of $64 \times 64 \times 256$. The ratio between the spatial and spectral dimensions has been defined regarding the typical source shape (Fig.~\ref{fig:bright_source_3d}) and constraints in both the training and prediction pipelines (Sects.~\ref{sec:training_example_gen} and ~\ref{sec:prediction_pipeline}). In the SDC2 data, and likely for any hyperspectral data as well, the signal distribution along the frequency dimension is very different from that of the two sky dimensions. Although all three axes are considered spatial for the box definition and matching criteria, we used anisotropic filters and stride configurations to account for the intrinsic differences in signal distribution across dimensions. Our best-performing SDC2 backbone, presented in Fig.~\ref{fig:network_architecture}, was obtained through an extensive search over the architecture and hyperparameter space using our SDC1 backbone as a starting point. The detailed configuration of our backbone architecture can be found in the training scripts archived at \href{https://doi.org/10.5281/zenodo.18403011}{10.5281/zenodo.18403011} (Sect.~\ref{sec:data_availability}). As we observed for SDC1 (Appendix B of Paper I), classical backbones developed for generic image datasets underperform when trained on the SDC2 task in comparison with a custom architecture. Using a pre-trained generic model as a starting point for training on the SDC2 task further degrades performance.

Despite being deeper with 23 convolutional layers, this new 3D architecture contains only $3.35$ million learnable weights, which is a quarter of what we had in our 17-layer SDC1 backbone. This mainly results from a reduction in the number of filters per layer, motivated by the reduced number of sources in the training catalog (1383, Sect.~\ref{sec:selection_function}). Still, the lesser morphological diversity in simulated HI sources compared to simulated continuum sources enables the network to achieve high detection and characterization accuracy with fewer learnable parameters. We note that the first layer convolves the frequency axis with filters spanning height channels, with a stride of two. This operation searches for large spectral structures and reduces the frequency axis dimension, compensating for the absence of frequency smoothing in the data preparation. Like our SDC1 backbone, the first few layers progressively reduce both the spatial dimensionality and the number of filters, acting primarily as a small denoising subnetwork tailored to this detection task. We also include the typical darknet sequence of alternating $3\times3$ convolutions with $1\times 1$ convolutions over a smaller number of filters \citep{paper:yolo_v2}. This forces the network to learn more compressed representations between spatial pattern extraction steps, reducing the number of learnable weights compared to stacking identically configured spatial convolutions. Here, these compression steps have a secondary purpose, which is to capture spectral patterns with 3D filters of size $1\times 1 \times 3$. This increases the receptive field along the spectral dimension faster than for the two sky dimensions, allowing the identification of extended patterns along the frequency axis. All layers except the last one use a leaky-ReLU activation with a 0.1 leakage factor. As with our SDC1 backbone, we have a group normalization layer between C20 and C21 with a group size of two \citep{paper:group_norm} and a 25\% dropout rate on the C21 layer \citep{paper:dropout}, which helps regularize the model.

Even though we could reduce the spectral dimension to obtain a 3D output grid of equal size across all dimensions, we observed that sampling the spectral domain over more grid elements grants better spectral localization. The total reduction factors for each dimension at the end of the network are then $8:8:16$, and each detection unit is responsible for an $8\times 8 \times 16$ input subvolume. The receptive field at the last layer is $66\times 66\times 294$, which corresponds to the input subvolume accessible to each detection unit for making a prediction. Removing layers anywhere in the final backbone measurably reduces the achievable score, while increasing the number of filters per layer leads to early overtraining. This likely indicates that the number of parameters is limited by the training sample size, and that network depth helps reach the required expressivity with fewer learnable parameters.

\subsection{YOLO-CIANNA SDC2 configuration}
\label{sec:method_setup}

The backbone architecture sets the output grid resolution, but the detection units themselves must be configured manually through a set of hyperparameters. In this section, all YOLO-CIANNA-related parameters follow the notations and definitions from Paper I. Some parameter values are provided for reproducibility and not necessarily discussed or commented on. The final configuration is the result of a thorough exploration of the corresponding hyperparameter space.

The typical volume source density in the SDC2 data is very low. Because our output grid already has a high resolution relative to the average source size, we observed that a single unit per grid element is sufficient to achieve the best score. If trained with more units, the detector often converges toward a solution where a single unit produces all the relevant detections. This is likely due to the small training sample and to the absence of clear and balanced subgroups in the source parameter space (Sect.~\ref{sec:parameter_space_correlations}). The background detection scaling factor is set to $\lambda_{\rm void}=0.01$.

As stated in Sect.~\ref{sec:yolo_cianna_desc}, our detection units represent objects as bounding boxes, which are essential to the training association function. As thoroughly discussed in Paper I, astronomical sources rarely exhibit the sharp and well-defined edges required to define bounding boxes. We therefore relied on a box proxy of size $S'$ in both sky axes and $D$ in the frequency axis for each source (Eq.~\ref{eq:source_size}). We rotated this box based on $PA$ and searched for the smallest box whose edges are aligned with the survey axis, and that contains all rotated vertices, thereby defining our final target box. The impact of the inclination angle, $i$, is already included in the definition of the source HI size. However, the resulting boxes remain small, which increases training difficulty as an unconstrained detector initially relies on fortuitous matches to identify the objects of interest. This problem would arise for every dataset, but it is more pronounced in 3D as positional errors across all dimensions are combined in the DIoU. To compensate, we increase the size of the target bounding boxes by 4 pixels in both sky axes and 10 channels in the frequency axis. We stress that this box definition has no impact on the HI size, $S$, and line width prediction, $w_{20}$, used in the scorer matching criterion, which are predicted as additional regression parameters. The corresponding target box is illustrated for a few sources in Fig~\ref{fig:match_reproj_set}. From this, we set our detection unit box size priors to $10\times 10\times 36$. We observed that small changes in the target box sizes can significantly affect the score, whereas changing the size of the prior has only a minimal effect.

Our single detection unit is configured to predict the box center coordinates ($x, y, z$), the box size ($w, h, d$), a detection probability, $P$, and an objectness score, $O$. To these mandatory elements, we add $f$, $S$, $w_{20}$, $PA$, and $i$, all expressed as individual linearly activated output vector elements constrained by a regression loss subpart. All parameters are normalized in a similar range to balance their loss contributions. $f$ is clipped in between [10,300] Jy\,Hz, log-scaled, and rescaled to [0,1]. $S$ is clipped in between [4,25] arcsec and linearly scaled to [0,1]. $w_{20}$ is clipped in between [100,700] km\,s$^{-1}$, log-scaled, and also rescaled to [0,1]. To avoid tasking the network with impossible predictions, we set $PA=180^\circ$ for all small sources with $S < 6$ arcsec (beam size is 7 arcsec). $PA$ is then converted from its [0,360] degree range to two independent parameters in the [-1,1] range through sin and cos functions, which works better than a direct prediction of the angle due to angular symmetries and periodicity. $i$ is converted from its [0,90] range to the [0,1] range through the cos function, which also flattens the otherwise inhomogeneous distribution of $i$. In summary, with six extra parameters, all targets and predictions are expressed in the form of a 14-element vector $\left< x, y, z, w, h, d, P, O, f, S, w_{20}, \sin{(PA)}, \cos{(PA)}, \cos{(i)}\right>$. The individual $\gamma^p$ scaling values are given in the top row of Table~\ref{table:hyper_param_list}.

The remaining method hyperparameters are listed in Tables~\ref{table:hyper_param_list} and ~\ref{table:scaling_param_list}. All fIoU thresholds are given for a DIoU-based association function. We note a few differences compared to the SDC1 setup, such as the absence of association refinement parameters, which are not required when using a single detection unit per grid element. Difficulty-flagging parameters are specified here, but discussed in the next section.

\begin{table}
\centering
\caption{\label{table:hyper_param_list} Detection hyperparameters.}
\begin{tabular}{ l c c c c c c}
\hline
\hline
 Param & $\gamma^{f}$ & $\gamma^{S}$ & $\gamma^{w_{20}}$ & $\gamma^{\sin{(PA)}}$ & $\gamma^{\cos{(PA)}}$ & $\gamma^{\cos{(i)}}$ \\
 Value & 3.0 & 2.0 & 2.0 & 1.0 & 1.0 & 2.0 \\
 \hline\\[-2.5ex]
 Param & $L^{{\rm fIoU}}_{\rm GBNB}$ & $L^{{\rm fIoU}}_{P}$ & $L^{{\rm fIoU}}_{O}$ & $L^{{\rm fIoU}}_{p}$ & $L^{{\rm fIoU}}_{\rm diff}$ & $L^{{\rm obj}}_{\rm diff}$\\
 Value & 0.4 & -0.1 & -0.1 & 0.4 & 0.1 & 0.2\\
\hline
\end{tabular}
\end{table}

\begin{table}[t]
\centering
\caption{\label{table:scaling_param_list} Scaling factors and limits of output subparts.}
\begin{tabular}{ l c c c c c}
\hline
\hline
& Pos. & Size & Prob. & Obj. & Param. \\
\hline
Activation & Sigm & Lin & Sigm & Sigm & Lin \\
$\lambda$  & 2.0 & 1.0 & 1.0 & 6.0 & 3.0 \\
Pre-activ. scaling &  0.5 &  0.5 &  0.2 &  0.5 &  0.5 \\
Pre-activ. max     &  8.0 &  1.8 &  8.0 &  8.0 &  1.2 \\
Pre-activ. min     & -8.0 & -1.4 & -8.0 & -8.0 & -0.2 \\
\hline
\end{tabular}
\end{table}

\subsection{Training example generation and augmentation}
\label{sec:training_example_gen}

Training examples were dynamically generated with augmentation from the normalized LDEV cube (Sect.~\ref{sec:cube_normalization}) and the selected LDEV truth source catalog (Sect.~\ref{sec:selection_function}). As discussed in Sect.~\ref{sec:cube_normalization}, we made the approximation that the frequency axis is an invariant spatial coordinate, meaning we trained our 3D detector to be insensitive to the input subvolume position on the frequency axis. We defined two ways of generating training subcubes. The first one, in 70\% of cases, extracts a cutout at a random position within the LDEV cube regardless of its content. Sources whose central coordinates are within the cutout are added to the target list. This mostly returns empty cubes, but it exposes the network to the full diversity of background signals and artifacts. The second one, in 30\% of cases, draws a random source from the selected truth sample and extracts a random cutout within the subregion of the LDEV cube where the selected source is guaranteed to be included. If other sources have their center in the same region, they are also added to the target list. With this approach, we ensure that a minimum proportion of examples contains at least one detectable source.

Extracted subcubes were augmented to increase the diversity of examples. We added masks on the edges of the subcubes to mimic the effect of added padding when applying the method to the edge of a large cube. Masks were added independently to each axis at a 20\% rate, always on a single side selected at random, and masking up to half the subcube in that axis. We also applied flips at a 50\% rate for each axis independently. Whereas this is a standard augmentation for sky axes, it is generally not used in the frequency dimension. It requires one to assume that the source morphology is symmetric along the frequency axis, which is a stronger approximation than treating the source frequency position as invariant. Still, enabling this augmentation improves the achievable score by a few percent, indicating that the increased example diversity outweighs the potential drawbacks of this approximation. Target $PA$ values are updated to account for flips when necessary.

The input resolution of the example subcubes was set to $64 \times 64 \times 256$. As we discussed in Sect.~3.8 of Paper I, when splitting a finite-sized volume, the choice of chunk size impacts the accessible diversity inside each chunk. The smaller the input volume, the higher the diversity, and the lower the chances of overtraining. A small context window also prevents the detector from inferring the input's position within the large cube, which could lead to overtraining. On the other hand, large input sizes are numerically efficient and reduce the proportion of the volume affected by edge effects and truncated content when making predictions (Sect.~\ref{sec:prediction_pipeline}). We observed that adding augmentations improves the results, indicating that the intrinsic diversity of the LDEV cube is a limiting factor at the selected input size. 

Our dynamic augmentation scheme does not fit the classical definition of an epoch. Instead, we grouped the generated examples into iterations over 1600 example subcubes. Due to the limited number of labeled sources in the LDEV catalog, splitting them into training and validation sets is impractical. Constituting a validation set large enough to be statistically relevant would remove too many sources from the training sample. We therefore chose to use all labeled sources from the LDEV catalog for training and relied on occasional scoring over the MAIN SDC2 cube to verify the absence of overtraining, serving as a validation and calibration dataset. During the challenge, the number of scoring was limited to prevent optimization against the hidden MAIN catalog. Yet, it was possible to score a few saved states for each trained model to test for blatant overfitting over the LDEV cube. In this paper, we used the scorer more extensively to provide a detailed analysis of our results. Still, it remains prohibitively costly to score more often than every few hundred iterations. Monitoring the training loss directly is not very useful, as most input examples lack detectable sources. Instead, we relied on a subset of cutouts centered on 800 randomly selected sources from the training catalog as a proxy validation dataset to monitor fine-grained loss evolution. Validation loss values are therefore only comparable within a given training.

\subsection{Training setup and bootstrapping}
\label{sec:training_setup}

\begin{table}[t]
\centering
\caption{\label{table:training_cat_content_bt} Training source count over multiple bootstrap steps.}
\begin{tabular}{ l c c c c c}
\hline
\hline
Step & Total & Match in & Match out & Similarity \\
\hline
Base &  1383 &      N/A &       N/A &        N/A \\
BT1  &  1573 &     1066 &       190 &    87.92\% \\
BT2  &  1565 &     1090 &       182 &    98.60\% \\
BT3  &  1574 &     1088 &       191 &    99.05\% \\ 
\hline
\end{tabular}
\end{table}

Training was performed with the \texttt{CIANNA} framework, which implements our YOLO-CIANNA method, using an RTX 6000 Ada with (91 FP16 TFLOPS). We used FP16C\_FP32A mixed-precision training, meaning that computations are done at a 16-bit floating point precision with 32-bit floating point accumulators \citep{paper:mixed_precision}. All data flowing through the network uses the reduced 16-bit precision. A 32-bit master copy of the weight is maintained and used to accumulate the weight updates. Training at this precision yields similar scores to those obtained from full 32-bit training. We set the batch size to 16, which offers a good balance between training efficiency and memory footprint. The learning rate starts at $1.2\times 10^{-5}$ and increases linearly over the first 32\,000 examples to $6 \times 10^{-4}$. It then decays exponentially with the number of training iterations at a rate of $5\times 10^{-4}$ toward a minimum of $1.2\times 10^{-5}$. The optimizer is a mini-batch stochastic gradient descent with momentum set at 0.7 and a weight decay of $5 \times 10^{-4}$. A typical training over 6000 iterations takes around 36 hours on an RTX 6000 Ada, with an average processing speed of 100 input cubes per second. Best score is generally achieved around iteration 4000.

We observed that small changes to the selection function described in Sect.~\ref{sec:selection_function} can significantly impact the final result, potentially even preventing training altogether. This is due to an oversimplified selection criterion that does not reflect the detector's actual capabilities. During training, sources that can be detected but are not included in the target list are considered false detections, forcing the model to lower the detection probability for all sources that present similar characteristics. Conversely, undetectable sources included in the target list force the model to increase the detection probability for background noise. These two types of mislabeling affect the fitting of the objectness curve, resulting in suboptimal detection performances. One possible solution is to rely on the detectors' self-assessed detection capabilities to refine the selection function. We achieved this by first training a BASE model from scratch with random weight initialization \citep[using Glorot normal,][]{paper:xavier_init} based on our handcrafted selection function. The best iteration was identified using the SDC2 scorer over the MAIN cube following the prediction pipeline. This model was then applied to the same LDEV cube used for training in order to produce a list of detections with predicted confidence scores. We extracted the 1600 detections with the highest objectness score. From this, we defined a new selection function with lower line-flux and volume brightness thresholds as
\begin{equation}
    \big(f \geq 100\big) \quad \textrm{or} \quad \big(f > 14 \; \text{and}\; V_b > 0.011 \big),
\label{eq:selection_function_bt}
\end{equation}
but with the additional constraint that the selected sources must match one of the detections produced by the previous model based on a DIoU threshold of 0.1. By doing so, we included fainter true sources that the first trained model considers detectable. However, most added sources are difficult to detect and would likely complicate the training startup of a new model. To compensate, we used the difficult flagging introduced in Appendix A.6.4 of Paper I and flagged all new sources that do not meet the original selection function. This allows the detector to learn from the obvious sources first and progressively refine its solution to include fainter ones. Sources that fulfill the original selection function but are not detected by the first model could be removed or flagged as difficult. However, we chose to keep them as regular targets, as the score achieved by this first model is already high, indicating no significant negative effect from these sources. Also, these missed sources might become detectable thanks to the added diversity provided by the new targets. The obtained refined target source catalog can then be used to retrain a bootstrap model.

This bootstrap process can be repeated as long as it improves the final score or until the content of the training sample converges. For the SDC2, we trained a BASE model and three successive bootstrap models: BT1, BT2, and BT3. Each model was used to refine the selection function for the next step. Instead of training every new model from scratch, we reused part of the previously trained model as a starting point. The last three layers of the loaded model were replaced with identically shaped layers but with randomly initialized weights. We show and discuss what happens when bootstrapping with a from-scratch model in Appendix~\ref{sec:from_scratch_bootstrap}. In table ~\ref{table:training_cat_content_bt}, we indicate the evolution of the number of training sources for each bootstrap step, the number of sources from the original selection function for which a match is found by the previous model, the number of fainter sources from the new selection function for which a match is found that are added to the training sample, and the fraction of the final sources that were already present in the training sample of the previous step. From this, we can see diminishing returns as we add bootstrap steps. The first step yields the greatest change in the training sample content, whereas there is only a 1\% difference (about ten sources) between BT2 and BT3 training samples. Overall, the training content is so similar across all steps after the first one that it is unlikely to result in significant score differences. This is mostly confirmed by the fact that the remaining score difference between our bootstrap models presented in Sect.~\ref{sec:results_score} is of the same order as the retraining variability on a given training dataset. The exact best-scoring step after the first one is mostly random. While only a few steps are sufficient in this specific context, the optimal number of bootstrap steps might be much larger on other datasets and strongly depend on the initial selection function and the exact strategy used to reinject new targets. These elements are further discussed in Appendix~\ref{sec:poor_select_bootstrap}, where we evaluate the effect of starting the training from a degraded initial selection function.

\subsection{Prediction pipeline}
\label{sec:prediction_pipeline}

\begin{figure*}
    \centering
    \includegraphics[width=1.0\hsize]{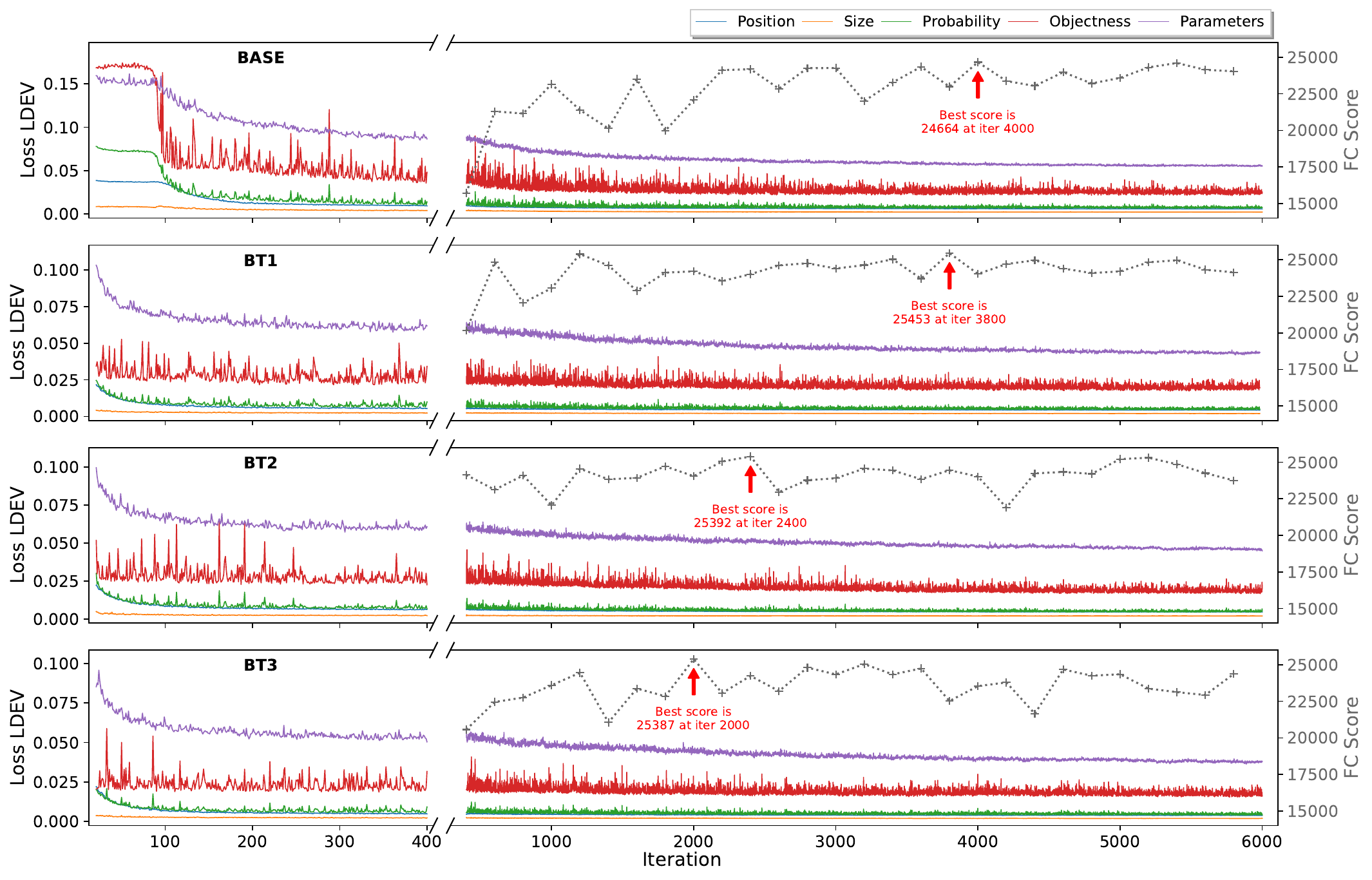}
    \caption{Evolution of the validation loss subparts (natural, Sect.~A.7 in Paper I) during training for our {four successive bootstrap models}. The first 20 iterations are omitted. Scores computed over the MAIN cube are indicated every 200 iterations after 400.}
    \label{fig:loss_curves}
\end{figure*}

To apply a trained detector to the whole SDC2 MAIN cube, we decomposed it into input regions along all three axes. To ensure that a source is fully contained in at least one input region, a minimum overlap of at least half the maximum source size is required in each dimension. In addition, the overlap must be a multiple of the backbone reduction factor to align subregions mapped by detection units from adjacent inputs, which is necessary for multiple detection filtering. We settled on an overlap of 8 pixels along the sky axes and of 32 channels along the frequency axis. This imposes constraints on the detector input size, which must be large enough to minimize the proportion of overlapping volume, as this directly affects the computational efficiency of the inference pipeline. The required wide frequency overlap was one of the motivations for having 256 frequency channels per input (Sects.~\ref{sec:network_backbone} and~\ref{sec:training_example_gen}).

Storing the MAIN cube in system memory and decomposing it dynamically is technically challenging. Thus, we rely on the cube slicing described in Sect.~\ref{sec:cube_normalization}. The detector is applied to all input regions of all subcubes, preserving only detections with an objectness score above 0.1. Multiple detections within each input are filtered through a first non-maximum suppression (NMS) using a 3D DIoU threshold of 0.1. Multiple detections from overlapping inputs are also removed using a secondary NMS with a threshold of -0.3. Detections are then transformed back to the SDC2 catalog format by converting 3D positions to RA, Dec, and central frequency, and by inverting the normalization applied to the predicted source parameters.

Classical computer vision tasks are often evaluated in a way that accounts for the detector's confidence score. In contrast, the SDC2 scorer expects a list of considered real detections. In practice, we can filter our prediction catalog using an objectness threshold deducted from the objectness histogram or by a rough estimate of the expected number of sources. The final threshold can be fine-tuned with a few calls to the scorer to optimize the result, which was our approach during the original challenge. However, to comprehensively analyze our results, we need to perform numerous comparable scoring on the MAIN cube, which requires an automated threshold search. This is achieved by scoring an unfiltered catalog, from which we extract the list of matches from the scorer products, including the individual score of each source. By binning our catalog by objectness and computing the average score in each bin, we can identify the threshold below which adding sources decreases the score. Since detection difficulty likely increases with frequency, we split the detected sources into 20 frequency bins and optimize independent thresholds for each bin. We acknowledge that this post-process threshold optimization technically breaks the challenge conditions, yielding a near 1\% improvement in maximum achievable score over a naive threshold selection. Still, this difference remains much smaller than the typical score difference observed between independent detection methods and allows us to compare all models produced with YOLO-CIANNA in their optimal selection regime, ensuring that the score difference between different setups or bootstrap steps are significative.

On an RTX 6000 Ada GPU, with inputs of $64\times 64\times 256$ and using FP16 compute resolution, we achieve a processing speed of 300 inputs per second, or 315 million voxels per second. The GPU memory footprint during inference is around 450 MB for a batch size of one and scales mostly linearly with the batch size, allowing the model to be used for prediction on entry-level hardware. The total processing time for the 450 GB datacube is approximately 30 minutes, evenly split between GPU processing and data loading from a high-speed NVMe SSD. This excludes the preprocessing time of the raw MAIN cube, which depends strongly on the available system memory. Post-processing time to filter the model's predictions is negligible.

\section{Results}
\label{sec:results}

\begin{table*}
\centering
\caption{\label{table:team_scores} SDC2 scores for source catalogs from different teams and methods.}
\begin{tabular}{ l c c c c c c}
 \hline
 \hline
 Team / method / model & $M_s$ (Score) & $N_{\rm det}$ & $N_{\textrm{match}}$ & $N_{\rm false}$ & Purity & $\bar{s}$ \\
 \hline
 \textit{Post-challenge results} & & & & & &\\
 \hline
 YOLO-CIANNA v1.0         & & & & & & \\
 \quad - BASE                                                   & \textbf{24\,664} & 36\,360 & 33\,404 & 2962 &          91.97\%  & 0.8269 \\
 \quad \textit{\small \quad $\hookrightarrow$ purity threshold} &         18\,459  & 22\,126 & 21\,927 &  199 &  \textbf{99.10\%} & 0.8509 \\
 \cmidrule{2-7}
 {\bf \quad - BT1}                                              & \textbf{25\,453} & 36\,971 & 34\,115 & 2862 &          92.28\%  & 0.8298 \\
 \quad \textit{\small \quad $\hookrightarrow$ purity threshold} &         19\,631  & 23\,505 & 23\,294 &  212 &  \textbf{99.10\%} & 0.8518 \\
 \cmidrule{2-7}
 \quad - BT2                                                    & \textbf{25\,392} & 36\,438 & 33\,785 & 2658 &           92.72\% & 0.8301 \\
 \quad \textit{\small \quad $\hookrightarrow$ purity threshold} &         19\,484  & 23\,354 & 23\,135 &  220 &  \textbf{99.06\%} & 0.8517 \\
 \cmidrule{2-7}
 \quad - {BT3}                                                    & \textbf{25\,387} & 36\,571 & 33\,791 & 2786 &           92.40\% & 0.8336 \\
 \quad \textit{\small \quad $\hookrightarrow$ purity threshold} &         19\,072  & 22\,752 & 22\,543 &  209 &  \textbf{99.08\%} & 0.8553 \\
 \hline
 \textit{Challenge results {\small ($M_s>10\,000$)}} & & & & & &\\
 \hline
 MINERVA v0.1$^*$                                               & \textbf{23\,254} & 32\,652 & 30\,841 & 1811 &           94.5\%  & 0.81   \\
 FORSKA-Sweden                                                  & \textbf{22\,489} & 33\,294 & 31\,507 & 1787 &           94.6\%  & 0.77   \\
 Team SOFIA                                                     & \textbf{16\,822} & 24\,923 & 23\,486 & 1437 &           94.2\%  & 0.78   \\
 NAOC-Tianlai                                                   & \textbf{14\,416} & 29\,151 & 26\,020 & 3131 &           89.3\%  & 0.67   \\
 HI-FRIENDS                                                     & \textbf{13\,903} & 21\,903 & 20\,828 & 1075 &           95.1\%  & 0.72   \\
 ... & ... & ... & ... & ... & ... & ... \\
 \hline
\end{tabular}
\caption*{\vspace{-0.1cm}\\ Note. The bold elements highlight the optimized metric for each result. \\ *Combination of the catalogs obtained with a prototype version of YOLO-CIANNA 3D and CHADHOC \citep{paper:sdc2}.\vspace{-0.5cm}}
\end{table*}

\subsection{MAIN cube scoring}
\label{sec:results_score}

We present results from four successive bootstrap trainings: BASE, BT1, BT2, and BT3. We show the evolution of all natural loss subparts on our proxy validation dataset as a function of the iteration for these four trainings in Fig.~\ref{fig:loss_curves}. We also represent the evolution of the score evaluated on the MAIN cube as a function of the iteration to confirm the absence of overtraining. We observe a plateau in the loss at the beginning of training for the BASE model. This plateau corresponds to a regime in which the network is still unable to identify the relevant signal for the task. Its length depends strongly on the method setup and backbone, with the worst-case scenario being an infinite plateau. The subspace of method parameters and architecture that enables the network to overcome this state is narrow and hard to find, highlighting the difficulty of the task. We observed that selecting only the few hundred brightest sources as targets reduces the plateau's length, but it also means fainter sources are never detected, resulting in a lower score. This is one of the motivations for using a bootstrap training approach, which allows the selection function to be refined over successive trainings. For a given selection function, there is no clear correlation between the plateau's length and the maximum achievable score when varying other parameters. Here, we optimized our method to reach the best score regardless of the plateau's length. BT1 to BT3 training losses do not exhibit the same initial loss plateau because these models rely on pre-trained layers from their respective previous bootstrap models, which had already learned the task's principles. We discuss how bootstrapping using from-scratch retraining of each model affects this initial plateau in Sect.~\ref{sec:from_scratch_bootstrap}. The observed short-term instability in both loss and score is caused by the low average source density, which leads to strong variation in typical source content between iterations. For each training, we selected the iteration with the highest score on the MAIN cube as the final model state. In principle, the best iteration should be determined based on an independent validation dataset distinct from both the training and final test datasets. However, as discussed in Sect.~\ref{sec:training_example_gen}, splitting the small LDEV catalog would further reduce an already small training source catalog.

Table~\ref{table:team_scores} presents a detailed score result comparison of the catalogs produced by our four models and of teams that scored above 10\,000 points in the original challenge. This table highlights the number of candidate detections, matches based on the scorer criteria, false positives, purity, and the average source score. A small description of each team's method and a detailed comparison of the produced catalogs are provided in the SDC2 summary paper \citep{paper:sdc2}. Here, we summarize a few additional observations regarding the original challenge scores. ML methods were used for all or part of the pipeline of several teams, notably by the two top-scoring teams and three of the four teams with the lowest submitted scores. This illustrates both the potential and the intrinsic difficulty of implementing such approaches for analyzing astronomical data. Also, most teams that applied denoising as a preprocessing failed to achieve a high score. This step likely removes signal from fainter sources, which are detectable only through their complex 3D signal distribution in a high noise environment. Although the first few layers of our backbone can act as a denoiser, the network learns which signal to compress or preserve regarding the specific SDC2 task. Most teams using classical approaches relied on the same software, \texttt{SoFiA} \citep{paper:sofia_2015, paper:sofia_2021}. Team SOFIA notably achieved third place in the challenge. The second best-scoring team, FORSKA-Sweden, used a 3D U-Net for source segmentation but also relied on \texttt{SoFiA} for a part of its source characterization. Still, the remaining score difference between the classical \texttt{SoFiA} method and the two top-scoring 3D-CNN-based approaches highlights a specific strength of this type of model in extracting complex 3D patterns in high noise contexts (Appendix~\ref{sec:local_reproject}). Interestingly, this contrasts with the results of \citet{paper:hi_source_finding_comparison}, which performed a comparative study of a few HI detection methods and observed that \texttt{SoFiA} and a V-net 3D-CNN-based method performed similarly. This also highlights the importance of the architecture, method setup, and quality of the training data. Although some teams published a detailed version of their pipeline after the challenge, such as FORSKA-Sweden \citep{paper:forska_sweden}, there is currently no published post-challenge score comparable to ours. To facilitate future comparisons, all models and catalogs presented in this paper are made publicly available (Sect.~\ref{sec:conclusion}).

Our new BASE model improves our top challenge score by $6.1\%$, and the BT1 model pushes this improvement to $9.5\%$, placing it $13.2\%$ higher than the second top score. The number of matched sources also increases by $10.6\%$ between our BT1 model and our top challenge score. We note that the typical variation in best achievable score over multiple training using the same setup is typically around $\pm0.6\%$. Additional bootstrap training steps after the first one produce no significant improvement, except for faster convergence, as illustrated by our BT2 and BT3 models. The BT1 model scoring higher is mostly due to chance, as the training sample for all three BT models does not change much (see Sect.~\ref{sec:training_setup}). BT2 and BT3 scores are within typical retraining variability using the same setup. This indicates that our base selection function is already a good starting point. In Appendix~\ref{sec:poor_select_bootstrap}, we show that the bootstrap approach can help reach similar scores when starting from a degraded selection function. We highlight that our MINERVA v0.1 score was achieved through a combination of two methods, a prototype version of YOLO-CIANNA, and a hybrid classical and ML method called CHADHOC \citep[Sect.~4.1.7 from][]{paper:sdc2}. At the time, the YOLO-CIANNA prototype was limited to $\sim$21\,000 points, meaning that we improved our method-specific score by around $20\%$. The distribution of predicted sources is mostly homogeneous over all three dimensions of the MAIN cube. We represent cutouts centered on a few matched sources in Appendix~\ref{sec:local_reproject}.

\begin{figure}
    \centering
    \includegraphics[width=1.0\hsize]{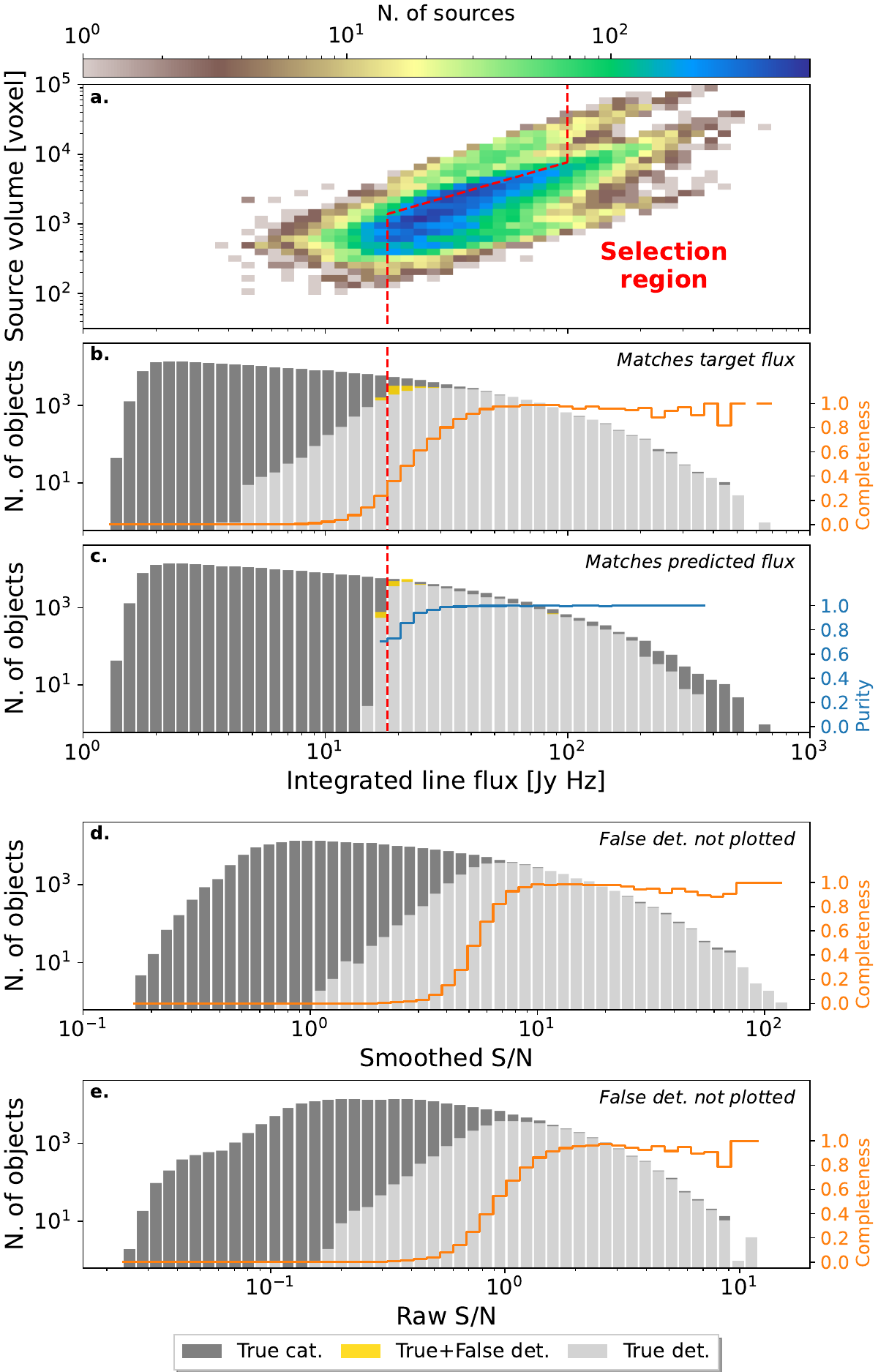}
    \caption{(a) Two-dimensional histogram of the MAIN cube detection matches target volume as a function of their target integrated line flux. The dashed red line represents the selection function as in Fig.~\ref{fig:selection_function}. (b) Histogram of the target line flux for the MAIN truth catalog and the matched sources. False positives are added based on their predicted flux. Completeness is overplotted for relevant bins. (c) Histogram of the predicted line flux for the matched predicted sources and false positives. The MAIN truth catalog is plotted in the background based on the target line flux. Purity is overplotted for relevant bins. (d) Histogram of the smoothed S/N for the MAIN truth catalog and the matched sources. Completeness is overplotted for relevant bins. (e) Identical to d but using the raw non-smoothed S/N.}
    \label{fig:pred_flux_dist}
\end{figure}

We represent histograms of the integrated line flux for both the matched and false detections in comparison to the complete MAIN truth catalog for BT1 in Fig.~\ref{fig:pred_flux_dist}. The first histogram is based on the target values and used to compute completeness as a function of the real line flux, whereas the second is based on the model predictions and used to calculate purity as a function of the predicted line flux. We also represent histograms of the target smoothed and non-smoothed S/N for the matched detections, as defined in Sect.~6.2 of \citet{paper:sdc2}. This S/N estimate was added to the MAIN truth catalog after the challenge and is therefore only used for analyzing the results. Finally, this figure shows a 2D histogram of the matched source volume against the target line flux. These plots show that false detections arise mainly in the low-flux regimes where completeness and purity drop smoothly. We also observe that the model correctly detects sources fainter than the limit imposed by our selection function, which is primarily due to our bootstrapping approach. Still, this ability to detect sources outside the training distribution is also observed in our BASE model's predictions, though to a lesser extent. These faint sources are likely rendered detectable by flux boosting from the local noise, as discussed in the next section. 

Our method produces a list of detections ordered by objectness confidence. Instead of selecting the threshold that maximizes the SDC2 score, we can search for a threshold that enforces a purity of at least $99\%$. We report the scores and properties of these purity-optimized catalogs for our four models in Table~\ref{table:team_scores}. The resulting catalogs still score 10 to $17\%$ higher than team SOFIA, the best non-ML approach. On the contrary, lowering the threshold results in more detection candidates. When applied to real observational data, these low-confidence detections could serve as follow-up targets for an observing program aimed at confirming some of them. The advantage of predicting a self-assessed confidence score is that a trained model samples the full detectability range. Therefore, as we did in Sect.~5.1.4 of Paper I, we computed an alternative metric similar to an average precision (AP). For this, we lowered our initial objectness prediction threshold to 0.05, sorted the predictions by objectness, and passed them to the SDC2 scorer to obtain the list of matches. We then computed the running precision (purity) and recall (completeness) over the catalog and converted them into a smoothed precision-recall curve, whose integral defines the SDC2-based AP. We report AP values of 18.35, 18.32, 18.34, and 18.25 for our BASE, BT1, BT2, and BT3 models, respectively. The total number of matches is close to 50\,000 for all models. To estimate the chances of random matches with a background source, we computed the AP of the BASE catalog after randomizing the 3D positions of all detections, yielding only 361 fortuitous matches. This indicates that detections in the low objectness regime contain a fair number of real detections, but the number of false detections is likely too high relative to the SDC2 scorer requirement. We also computed a purely geometric DIoU-based AP score for all models, which is presented in Appendix~\ref{sec:det_only_performances}.

\subsection{Source characterization quality}
\label{section:source_characterization}

\begin{figure*}
    \centering
    \includegraphics[width=1.0\hsize]{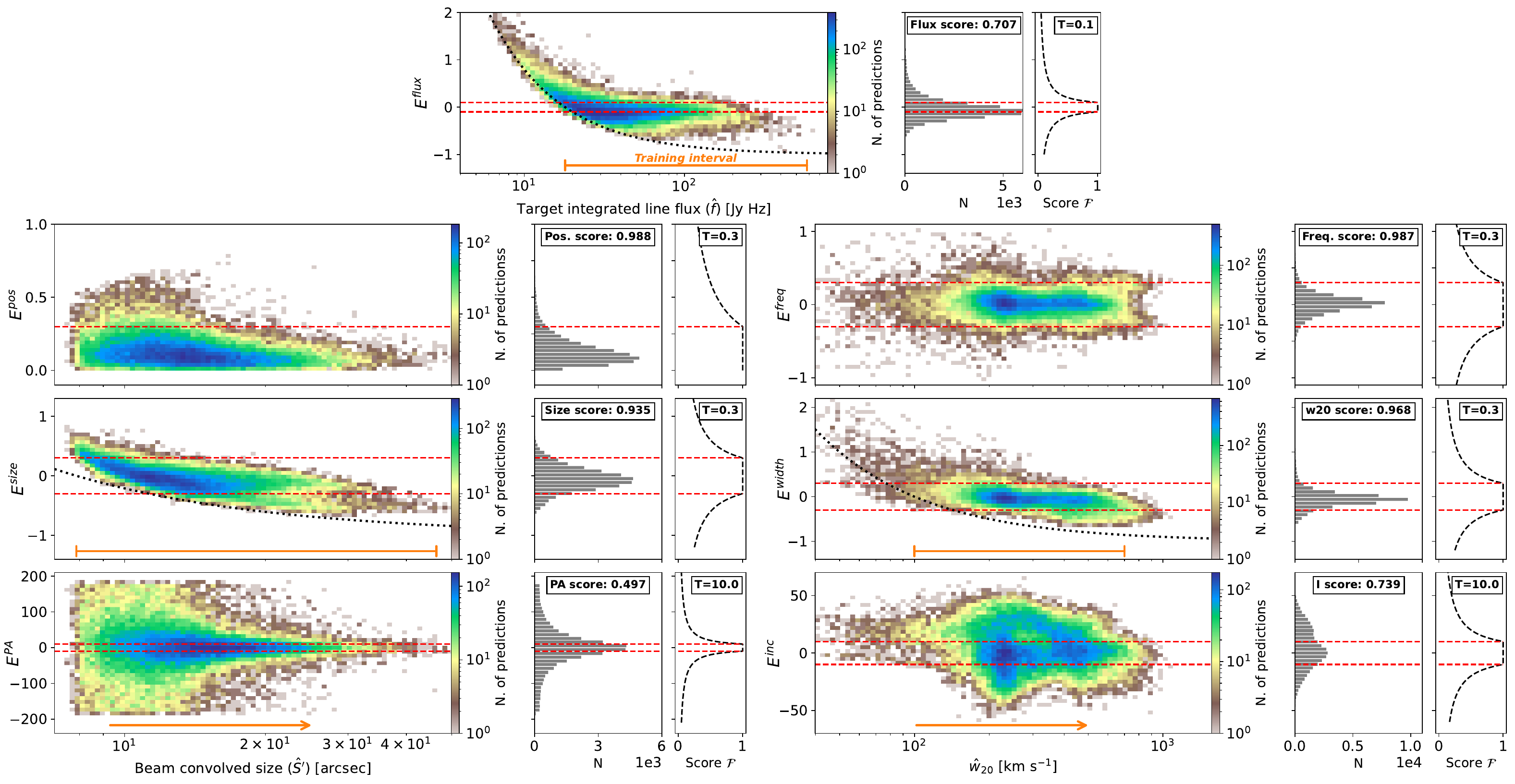}
    \caption{Subpart error distributions as 2D histograms against relevant target quantities or scales. For each subpart, the corresponding 1D error histogram and the associated score response functions are provided. The dashed red lines indicate the error region where the score saturates at 1. Orange lines and arrows indicate the range of values from the training sample when relevant. Dotted black lines represent the error obtained when predicting the minimum value from the training sample across the full comparison scale.}
    \label{fig:all_param_pred}
\end{figure*}

Our BT1 model achieves an average characterization score of $\bar{s}=0.8298$, improving our original challenge submission by 0.02, which was already the highest characterization score among all teams. When multiplied by the approximate 34\,000 matches, this results in a direct increase of about 680 points. However, an average characterization score is not suitable for comparing methods with varying detection performances. This penalizes strong detectors that successfully detect many faint sources, which are likely also difficult to characterize. This is illustrated by our alternative purity-optimized catalogs that all achieve $\bar{s}>0.85$ by selecting only the most obvious detections. A better approach would be to evaluate the average characterization score for catalog intersections for each pair of methods to compare, so it is computed on the same list of sources. However, to date, the catalogs submitted by teams that participated in the original challenge have not been publicly distributed. This is one of the motivations for distributing all our SDC2 catalogs, which should enable future comparisons (Sect.~\ref{sec:discussion}). 

We also note that the scorer uses relative errors, which lead to an asymmetry between positive and negative errors, with the positive side unbounded and the negative side limited to minus one. When combined with symmetric score response functions, catalogs that systematically underestimate parameters achieve higher scores than those with a zero-centered error distribution. This bias is inherited from the SDC1 scorer design as discussed in Sect.~5.1.3 of Paper I.

To thoroughly evaluate the characterization accuracy of our method, we rely on the detailed information exposed by the scorer code. For each match, it provides the predicted and target source properties, the matching distance, all error subpart values $E$ from Table~\ref{table:error_functions}, and the final source score recombined by Eq.~\ref{eq:score_response_fct}. From this, we draw individual error subpart histograms and compare the resulting error distribution with the associated score response function, both shown in gray in Fig.~\ref{fig:all_param_pred}. For each error subpart, we use dashed red lines to indicate the region of the distribution where the score is saturated. The more the error distribution is contained in the corresponding area, the higher the average subpart score. For each error subpart, we identified an appropriate source target property against which it is relevant to represent the error distribution. For example, we represent the flux error as a function of the target's integrated line flux, which is natural considering that $E^{\rm flux}$ is simply a relative flux error. The sky position, HI size, and PA are represented against the target beam-convolved HI size. The frequency position, line width, and inclination are represented against the target line width. All these distributions are represented in the form of 2D histograms, presented alongside their 1D counterparts in Fig.~\ref{fig:all_param_pred}.

From this figure, we observe that faint sources exhibit a strong positive error. This results from multiple effects. Firstly, the cube noise imposes a hard limit on the minimum predictable flux. Secondly, the random fluctuation of the noise can either boost or reduce the apparent flux. Sources with per-voxel flux near the noise level can only be detected when the noise contribution is positive, thereby acting as a selection effect of sources with overestimated flux. This effect is reinforced by the detector's inability to predict a flux below the minimum target value seen during training. This is illustrated by the dashed black line over the 2D $E^{\rm flux}$ error distribution representing the error that would arise by systematically predicting the minimum training flux. Considering that the scorer includes the flux error in its matching criteria (Eq.~\ref{eq:match_tot}), properly detected faint sources might be rejected based on high flux error, even though we reached the best achievable flux prediction for the noise level. Although the scorer does not expose the $D_{\rm tot}$ value for rejected sources, we observe that the faintest matches already have values close to the $D_{\rm tot} \geq 5$ rejection limit, suggesting that some of our false positives might fall in this category. On the contrary, the flux of the brightest sources is underestimated, which is mainly caused by the scarcity of bright examples in the training sample.

We observe similar behavior in both source size and line width error distributions, which also result from observational limits, specifically the beam size and channel resolution. These subparts have higher average scores due to less strict error thresholds. Both the sky position and the central frequency appear well characterized, as would likely be the case for any catalog of matches, since poorly positioned detections have already been rejected by matching criteria stricter than the associated subscores. We note that despite their resemblance, $E^{\rm pos}$ and $D_{\rm pos}$ are two different quantities, as they rely on different reference sizes ($S' \neq S''$), which explains why the position error distribution does not expand up to the matching rejection limit of one. Finally, the position angle is well characterized for sources with sufficient HI size but is predicted at random for small sources, which is also a consequence of the observational limit imposed by the beam size. The inclination angle achieves a better score. However, the substructures visible in the distribution as a function of the line width suggest that this angle might be inferred from correlated parameters rather than directly measured from the cube. We confirm that the network learns parameter correlations to some degree in Appendix~\ref{sec:parameter_space_correlations} by representing the source distribution for a selected set of parameter couples.

\section{Discussion}
\label{sec:discussion}

This second edition of the SKAO SDCs improves the challenge format in several ways compared to the SDC1. The simulated cube includes more physical and instrumental effects, improving its realism, which is also a strength of the subsequent SDC editions \citep[see the SDC3a,][]{paper:sdc3a}. The LDEV cube is a proper representation of the MAIN cube, allowing one to train supervised ML methods without requiring representativity corrections like we implemented in Paper I. Still, the use of a similar scoring method to the SDC1 results in the same bias and limits, such as strong coupling between detection and characterization performance evaluation, or the absence of a confidence score weighting like in most computer-vision challenges. We also acknowledge that using simulated data provides an unrealistic advantage for training supervised ML methods compared to real observations. All sources are known and perfectly characterized, allowing for the definition of unambiguous targets, which would not be the case with observational labeled catalogs.

The results obtained with our YOLO-CIANNA 3D method, combined with the fact that the two top-scoring teams from the SDC2 utilized 3D CNNs, demonstrate that this type of structure offers significant advantages for processing noisy hyperspectral data. However, as discussed in Paper I, this requires one to build dedicated approaches that account for the specificities of astronomical data at the heart of their design. For this study, this was facilitated by the low-level nature of our \texttt{CIANNA} framework. Nevertheless, several aspects of our method could be improved. In terms of backbone architecture, \texttt{CIANNA} V-1.0 does not implement residual and skip connection layers \citep{paper:residual_connect}, which limit the maximum network depth. Such layers have been added to the experimental build of \texttt{CIANNA}, which should be publicly released on a short timescale. Unlike some versions of YOLO \citep{paper:yolo_v3}, we did not split predictions across scales, which is also related to current architectural limitations. In addition, having a fixed number of detection units on a finite-sized grid constrains both the positioning resolution and the density of detectable sources. This limit could likely be overcome by adding attention layers \citep{paper:transformer, paper:detr, paper:vit_yolo} and modifying the definition of the output layer. We are also exploring pre-training of our backbone using the denoising diffusion probabilistic model \citep[DDPM,][]{paper:ddpm} formalism. This could be used to pre-train a model over the MAIN cube without requiring any source targets. Finally, we could inject context information, such as the instrumental setup, at multiple stages of the backbone to guide the source extraction \citep{paper:vilbert}. This would enable us to train generic detection models compatible with multiple setups or instruments. 

The natural next step after achieving state-of-the-art results on simulated data is to generalize our method application to observational data from SKA precursors such as the LADUMA \citep[Looking at the Distant Universe with the MeerKAT Array,][]{paper:laduma} and WALLABY \citep[Widefield ASKAP L-band Legacy All-sky Blind surveY,][]{paper:wallaby} surveys. Our method could also be generalized to detect line emission in hyperspectral surveys covering a different frequency domain, such as MUSE. However, it requires a large set of labeled examples to build the training and validation sets. Although simulations can generate large example datasets for pre-training highly parametric statistical models, they often lose accuracy when generalized to real observations. A complementary training phase on real labeled data is almost always necessary, and the inference pipeline must necessarily be calibrated on a labeled real validation dataset. For this, we would combine labels from classical detection methods, crowdsourced science, or existing catalogs from other instruments. The supervised nature of our method implies that the trained detector would likely inherit uncertainties and biases from the methods used to define the targets. Combining independent labeling methods might help in this regard. In all cases, the bootstrap training strategy can be used to train a model from an imperfect training sample to identify new candidates, which can then be confirmed using other labeling approaches or new observational programs, iteratively improving detection sensitivity. Preparatory work indicates that direct porting of our SDC2-trained models generalizes well to preliminary LADUMA data after a simple rescaling, producing candidate detections compatible with catalogs obtained with \texttt{SOFIA}. Specific retraining for this survey is currently being explored.

We conclude by updating the numerical environmental footprint estimated in Paper I, which already accounts for most of the impact associated with the YOLO-CIANNA method development and testing for 2D and 3D datasets. For this SDC2 study, we estimate the specific additional computational time to 4000 GPU hours, which includes the 1000 hours granted to MINERVA on Jean-Zay during the original challenge. Using the same estimates of 0.5 kW of average system power consumption when training with a single GPU, and France's carbon intensity estimated by the ADEME for 2022 at 52 g${\rm CO_2\text{-e}/kWh}$, the added impact is about 104 kg of ${\rm CO_2\text{-e}}$.

\section{Conclusion}
\label{sec:conclusion}

In this paper, we generalize our YOLO-CIANNA deep learning and characterization method to 3D hyperspectral data from the SKAO SDC2. We detail the required modifications to the box definition, the position and size regression loss subparts, and the match association metric. We present a backbone optimized for HI line-emission detection and describe the construction of our training sample and our training and inference pipelines.

Our results show that 3D CNN structures efficiently extract complex 3D patterns specific to HI sources and achieve best-in-class source characterization. However, we highlight that implementing such a method on complex astronomical data is challenging in practice, and that only a narrow region of the parameter and architecture space enables successful training. We also introduce a bootstrap training strategy that relies on the self-assessed detection capability of successively trained models to refine the selection function used to build the training sample.

These state-of-the-art results are encouraging regarding the generalization of the method to observed surveys from SKA precursor instruments, with preparatory work currently underway for the LADUMA and WALLABY surveys. Although 3D CNN approaches should not be considered drop-in replacements for classical analysis tools, which often offer greater interpretability and consistency, our results demonstrate that deep learning detectors can efficiently produce complete and pure first-guest catalogs from large hyperspectral datasets.

\section{Data availability}
\label{sec:data_availability}

Example scripts and notebooks are provided in the \texttt{CIANNA} git repository to help reproduce our results. We also publish all training and inference codes, trained models, and detection catalogs presented in this paper at \href{https://doi.org/10.5281/zenodo.18403011}{10.5281/zenodo.18403011}, along with ancillary models trained over the whole MAIN cube to serve as pre-training states for transfer learning toward other data or surveys (Appendix~\ref{sec:alt_models}).

\begin{acknowledgements}

We acknowledge the support granted to the MINERVA project by the Paris Observatory, including material and financial support. We acknowledge the Paris Observatory DIO shared computing resources and their handling of the MINERVA dedicated compute resources. We acknowledge the computing resources from the GENCI (Jean-Zay GPU partition) provided to the MINERVA team in the context of the SDC2. DC acknowledges the MINERVA project for funding a two-year post-doctoral contract. DC acknowledges its PSL fellowship as part of the MCA-IA program and in collaboration with PR[AI]RIE-PSAI. We highlight the great work SKAO has done in organizing all the SDCs and thank them for allowing post-challenge access to the datasets.

\end{acknowledgements}

\bibliographystyle{aa}
\bibliography{aa57257-25.bib}

\begin{appendix}
\onecolumn

\section{Source parameter space correlations}
\label{sec:parameter_space_correlations}

\begin{figure*}[h]
    \centering
    \includegraphics[width=1.0\hsize]{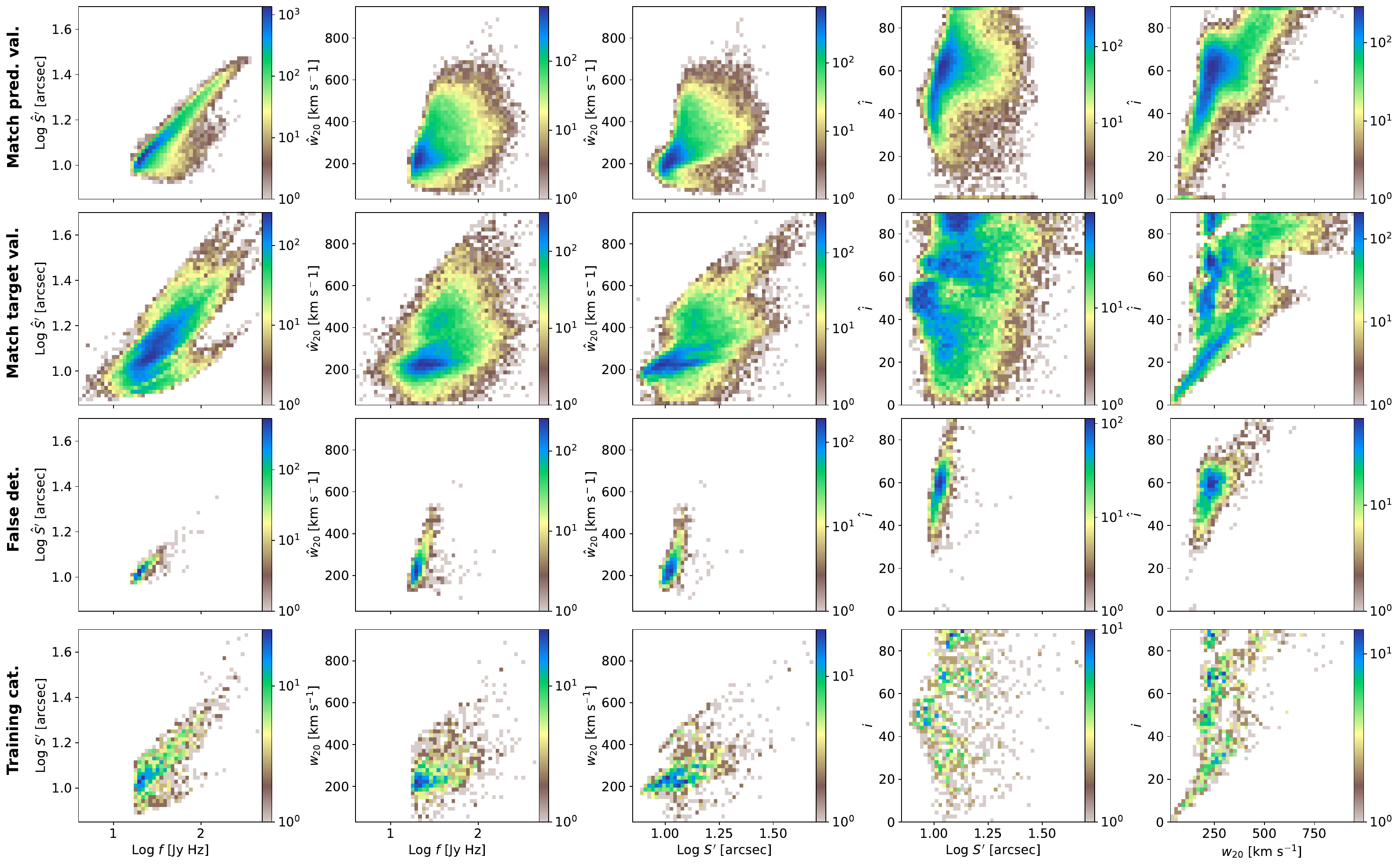}
    \caption{Two-dimensional histograms of selected source parameter pairs. Rows correspond to different source catalogs: the \textit{first} and \textit{second} rows represent matched sources from the MAIN from BT1 using the predicted and target source values, respectively; the \textit{third} row represents false detections; the \textit{fourth} row represents the selected training catalog from the LDEV cube.} 
    \label{fig:multi_correl_data}
\end{figure*}

Both the MAIN and LDEV truth catalogs exhibit correlations between source parameters. As we observed in Sect.~\ref{section:source_characterization}, the inclination angle error distribution exhibits substructures, suggesting that the network likely infers this parameter from other source properties. We identified parameter pairs that exhibit correlations and present them in the form of 2D histograms for all matched sources in our BT1 catalog in Fig.~\ref{fig:multi_correl_data}. To determine whether our detector reconstructs these correlations, we show the distributions for both the predicted and associated target values. We also display histograms for candidate detections rejected by the scorer's matching criteria and for the training sample to determine if the parameter space coverage is sufficient for these correlations to be clearly identified.

We observe a linear correlation between the flux and the HI source size. We also observe substructured distributions in both histograms that involve the inclination angle. Our detector fails to capture most fine-grained substructures, reducing them to mostly linear relations, which is likely due to the sparse sampling of the parameter space by the training sample. Still, the observed relations confirm that the model learns to correlate source properties, allowing it to infer some parameters from others that are easier to evaluate. Finally, we observe that most false detections are faint sources with a small HI size and line width that fall in the lower half of the distribution. Although not shown here, we note that false detections are uniformly distributed across sky position and central frequency.

\section{Local reprojection effect on detectability}
\label{sec:local_reproject}

\begin{figure*}[h]
    \centering
    \includegraphics[width=0.92\hsize]{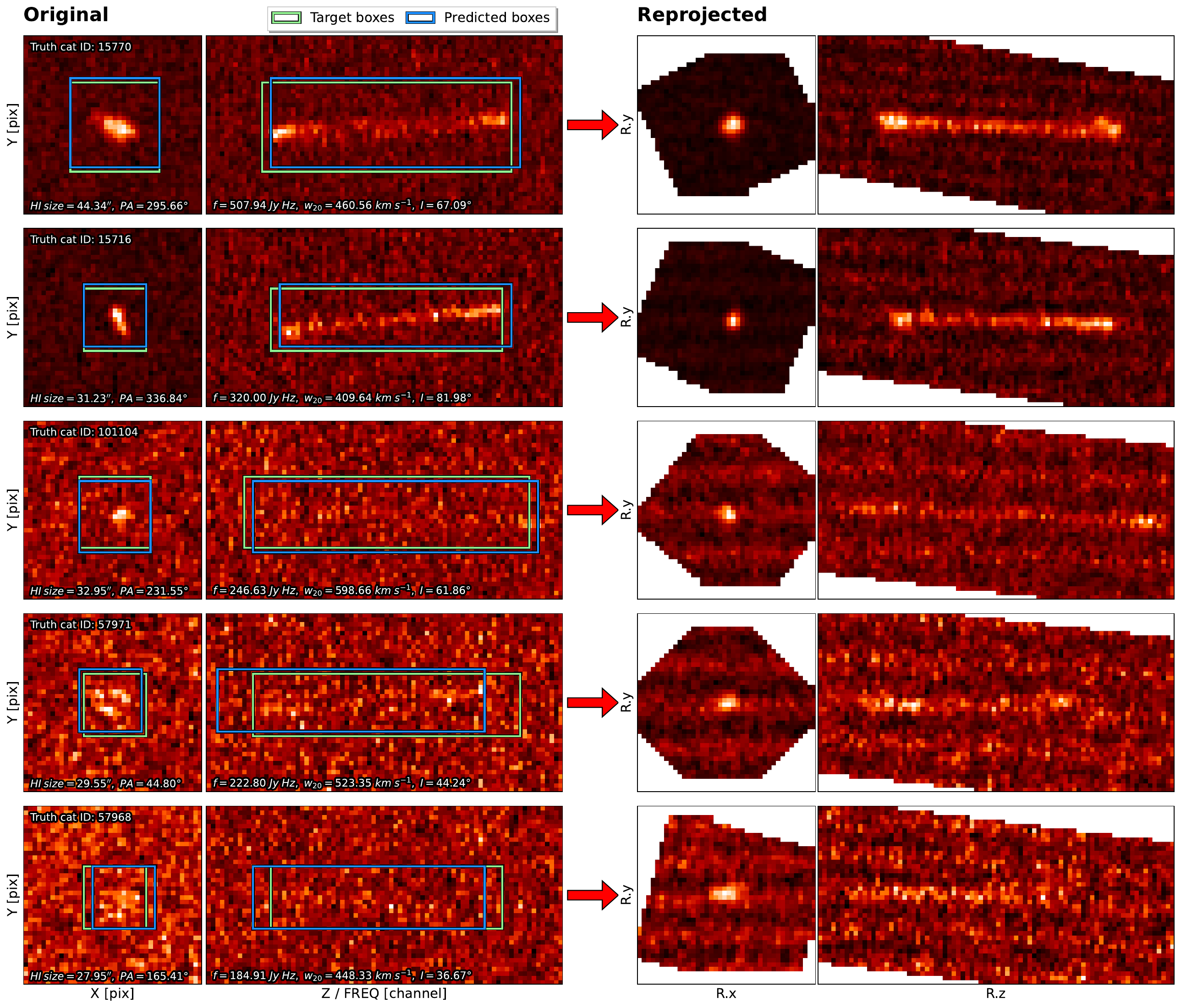}
    \caption{Signal projection for a subset of matched MAIN-cube detections. Each row corresponds to a source. The \textit{left} column is obtained from cutouts centered on the source in the original MAIN cube, whereas the \textit{right} column is obtained after reprojecting the cutout based on each source's $PA$ and $i$. For each cutout, we present two 2D projections, one in the plane of the sky by averaging over the frequency axis, and the second in a sky-frequency plane by averaging over one sky axis. Images are in arbitrary units inversely proportional to the normalized voxel intensity. The colormap is adjusted to maximize the per-image contrast. The truth catalog \texttt{id} and the target source properties are indicated for reference. Boxes are overplotted on the \textit{left} column for each source and represent the target box in green and the corresponding detector's prediction in blue.}
    \label{fig:match_reproj_set}
\end{figure*}

Like most participating teams, upon getting access to the SDC2 data, our first action was to explore the cubes visually. This approach allowed us to identify only a handful of very bright sources. Smoothing the LDEV cube in the frequency axis can increase the number of visually identified sources up to a hundred ($\sim1\%$ of the LDEV catalog). Directly visualizing the source signal in 3D, as we did with Fig.~\ref{fig:bright_source_3d}, is only possible for the brightest sources. Fainter sources can be visualized by taking a cutout centered on the source coordinate and projecting their signal along one of the cube axes based on the source extension, which boosts the S/N. This per-source approach obviously outperforms global cube smoothing in enhancing the individual source signal. However, it requires prior knowledge of each source's position and extension, which defeats the purpose of a detection and characterization task. Still, this test confirms that around 10\% of sources have sufficiently high signal density to be detectable, which remains lower than the capabilities of our YOLO-CIANNA detector.

Here, we aim to identify potential explanations for the robust detection capability of our method on this task. We hypothesize that 3D CNNs can learn to recombine the input signal in a way that enables them to dynamically explore multiple 3D reprojections and various source extensions based on the input context. To test this, we can identify the source main signal plane using the known truth PA and inclination angles, and reproject the cube to align with it. We do this for a few successfully detected sources from the MAIN cube and present the results in Fig.~\ref{fig:match_reproj_set}. To better illustrate the effect of this per-source reprojection, we select sources covering an extensive range of S/N and compare their 2D projections in the original cube and in the reprojected cutouts. The first source is the same as the one represented in Fig.~\ref{fig:bright_source_3d}. Regular horizontal lines visible in the reprojected space are artifacts of the cube reprojection method we used. Boxes corresponding to our target definition and the associated network prediction are overplotted on the projections from the original cube.

As expected, reprojecting based on known source characteristics boosts the signal. Any classical detection methods applied directly to this reprojected space would likely achieve similar performance to that of our detector. The strength of our method lies in its ability to identify such transformations automatically on the fly for each source. The fully convolutional 3D translational-equivariant structure of our backbone implies that our detector behaves as if the same sub-detector were simultaneously applied to every subregion mapped by our network, as defined by the output grid. Because this sub-detector is built from stacks of 3D convolutions, it can simultaneously scan for a wide range of 3D patterns that can typically emulate the expected projections. Achieving the same result with a classical method would require one to follow a sliding window approach, testing all possible positions in the 3D cubes and multiple detector configurations to account for the variety of suitable reprojections, which would be computationally prohibitive.

\FloatBarrier

\section{Detection-only task performances}
\label{sec:det_only_performances}

In Sect.~\ref{section:source_characterization}, we discussed how the SDC2 scorer correlates the detection and source characterization and how it affects the evaluation of the detector's capability. Here, we evaluate whether merging detection and characterization into a single model compromises pure detection performance. This question arises because the YOLO-CIANNA loss combines multiple objectives that are balanced through scaling factors (Eq.~14 in Paper I), which may compete against each other. This question is addressed for the specific case of the SDC1 in Appendix~D of Paper I, where a DIoU-based AP score is used to evaluate the purely geometric detection capabilities of our method across different training setups. It concludes that our combined detection and characterization model performs systematically better than equivalent models for which the parameter prediction subloss is deactivated. Although a naive combined loss could slightly reduce detection performance, the cascading loss design in YOLO-CIANNA ensures that the characterization target is used only for sources that are already confidently detected, thereby improving both detection and characterization results. These observations align with our general hypothesis that simultaneously predicting correlated variables often yields better results than making independent predictions.

In this section, we present the results of a similar verification for the SDC2, computed using purely geometric 3D DIoU-based AP scores at different thresholds for our four reference models from Sect.~\ref{sec:results_score}, to which we added a model trained with parameter prediction turned off (No-param). This new model was trained using the BASE model configuration and with our default selection function (Sect. ~\ref{sec:selection_function}). The absence of parameter predictions prevents bootstrapping, as it relies on the scorer matching criteria. We compute AP scores for DIoU thresholds of -0.3, 0.1, and 0.5, which correspond to the DIoU limits used in the second and first NMS steps of the prediction pipeline, and to a strict high-quality match criterion, respectively. The absence of source characteristic predictions also prevents us from using the scorer to identify the best iteration. We therefore rely directly on the geometric AP scores to select the best iteration for the No-param model, while keeping the best iteration based on the SDC2 scorer for all other models. We note that this choice might give the No-param model a small unfair advantage regarding AP scores. We present the purely geometric AP scores in Table~\ref{table:no_param_refine}, along with the scorer-based AP for our four reference models. The model without source characterization matches our BASE model's geometric AP scores at the first two thresholds but falls behind for the stricter threshold. It appears that the added parameter prediction improves source positioning and sizing but has no substantial impact on source detectability. Still, we can confirm that simultaneous detection and characterization have no adverse effect on pure detection accuracy using our YOLO-CIANNA method on the SDC2.

\begin{table*}[h]
\centering
\caption{\label{table:no_param_refine} Detection-only score metrics with and without the extra-parameter prediction enabled during training.}
\begin{tabular}{ l c c c c c c | c c c}
 \hline
 \hline
 Model &$\textrm{AP}_{-0.3}$ & $\textrm{AP}_{0.1}$ & $\textrm{AP}_{0.5}$ & $N^{m}_{-0.3}$ & $N^{m}_{0.1}$ & $N^{m}_{0.5}$ & $\textrm{AP}_{SDC2}$ & $N^{m}_{SDC2}$ & $\bar{s}$\\
 \hline
With param. BASE & 19.20 & 16.72 & 9.01 & 54\,022 & 46\,040 & 27\,476 & 18.35 & 51\,520 & 0.7816 \\
With param. BT1  & 19.34 & 16.99 & 9.70 & 52\,524 & 45\,323 & 28\,567 & 18.32 & 49\,306 & 0.7908 \\
With param. BT2  & 19.33 & 16.87 & 9.29 & 52\,768 & 45\,302 & 27\,806 & 18.34 & 49\,579 & 0.7903 \\
{With param. BT3}  & 19.26 & 16.89 & 9.64 & 52\,672 & 45\,282 & 28\,407 & 18.25 & 49\,437 & 0.7910 \\
No-param.        & 19.24 & 16.69 & 8.18 & 55\,996 & 46\,997 & 26\,807 & \textrm{N/A} & \textrm{N/A} & \textrm{N/A}\\
\hline
\end{tabular}
\end{table*}

\section{Bootstrap with from-scratch training}
\label{sec:from_scratch_bootstrap}

\begin{figure*}[h]
    \centering
    \includegraphics[width=1.0\hsize]{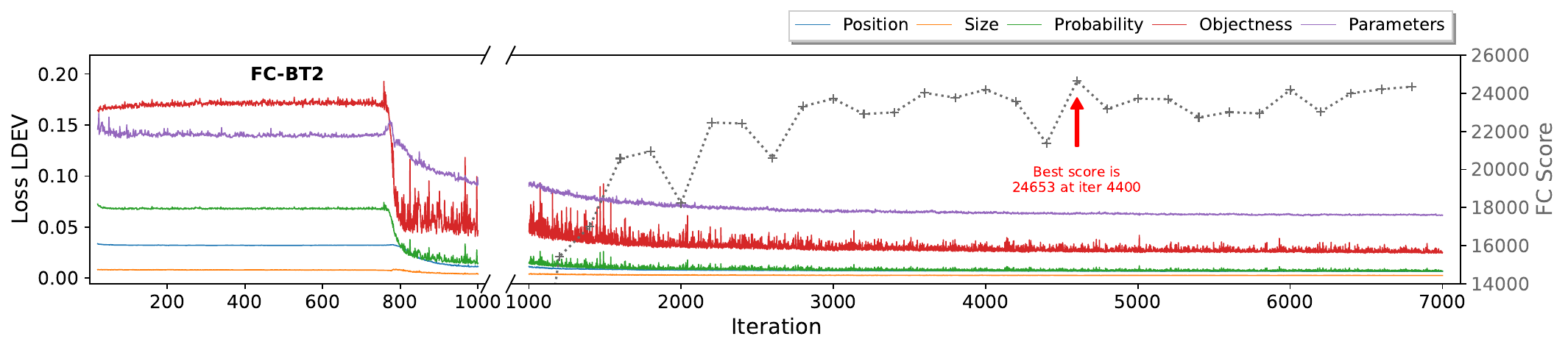}
    \caption{Evolution of the validation loss subparts (natural, Sect.~A.7 in Paper I) during training for our FS-BT2 model. The first 20 iterations are omitted. Scores computed over the MAIN cube are indicated every 200 iterations starting at 1000.}
    \label{fig:loss_curves_FS-BT2}
\end{figure*}

In Sect.~\ref{sec:training_setup}, we stated that during bootstrap, we reuse part of the trained layers from the previous model at each step, simply reinitializing the last three layers with random weights. This approach allows us to start from a model already capable of identifying obvious objects, removing the initial training plateau observed in the BASE model's validation loss in Fig.~\ref{fig:loss_curves}. The training can then focus on identifying fainter objects added to the training sample based on previous model detections. Here, we provide a reference point for what happens when starting with a fully randomized set of weights for all layers (from-scratch training) during bootstrapping. To exacerbate the effects, we retrained the equivalent of our BT2 model, which corresponds to a second bootstrap step, creating an FS-BT2 model. We represent the evolution of the validation loss as a function of the training iteration in Fig.~\ref{fig:loss_curves_FS-BT2}.

We observe an initial loss plateau for around 800 iterations, indicating that the model has difficulties in understanding the task and detecting even the brightest objects. This is mostly due to the addition of fainter targets to the training sample, which reduces the proportion of cases with an obvious object and adds training noise by requiring the model to detect sources in regions dominated by noise. This is typically the regime that the starting selection function tries to avoid. While flagging added sources as difficult helps mitigate the effect, it still affects training, especially given the reduction in the proportion of easy cases presented to the model due to our training example generation strategy (see Sect.~\ref{sec:training_example_gen}). We also observe that the maximum score achieved by FS-BT2 (24\,653 points with 33\,228 matches and 2606 false detections) is measurably lower than that of our BT2 model and closer to that of our BASE model, and that it is obtained at a later iteration than BT2. In theory, as long as the model can break out of the initial plateau, it should converge to a score similar to that of BT2, since they share the same training sample. We note that the average source score is lower than that of all other models, with $\bar{s}=0.8201$, but this accounts for only half of the score difference ($\approx$300 points). The rest most likely comes from a specific dynamic of the training: some YOLO-CIANNA loss subparts likely start overtraining (likely the parameter prediction) before others converge, due to the delayed beginning of global convergence. Such behavior could be addressed by adjusting the loss subpart scaling or other hyperparameters, but this would create an inconsistency in the training setup across successive bootstrap steps. From these observations, we consider our default approach of reusing part of the previously trained model more efficient than systematic from-scratch retraining.

\FloatBarrier

\section{Bootstrap from a degraded selection function}
\label{sec:poor_select_bootstrap}

We show in Sect.~\ref{sec:results_score} that our bootstrap training approach introduced in Sect.~\ref{sec:training_setup} significantly improves the score of our BASE model. However, the improvement may appear modest given the increased training time and pipeline complexity. We hypothesize that our initial selection function is already near optimal, and that the bootstrap method would allow us to achieve a similar final score if we started from a poorer selection. To test this, we defined a degraded selection function as
\begin{equation}
    \big(f \geq 100\big) \quad \textrm{or} \quad \big(f > 20 \; \text{and}\; V_b > 0.016 \big)
\label{eq:poor_selection_function}
\end{equation}
selecting 984 sources as detectable, which represents approximately 70\% of the sources used for our original BASE training. Selecting only brighter sources should accelerate model training convergence, but it is likely to result in a model that cannot detect fainter sources. We then applied two successive bootstrap training steps, yielding three trained models: B-BASE, B-BT1, and B-BT2. We summarize the results in Table~\ref{table:poor_select_scores}. 

The B-BASE model catalog scores 6\% below our BASE model with a 4\% drop in matches and a lower average source score, which is consistent with a reduced training parameter-space coverage. After a single bootstrap step, the B-BT1 model already achieves a score similar to that of our BASE model. The second bootstrap further improves the score by placing B-BT2 in between our BASE and BT1 models, confirming our hypothesis. Here, we followed the approach described in Sect.~\ref{sec:training_setup} to add new sources to the training sample, thereby limiting the gains from the degraded selection function. In particular, all new sources are flagged as difficult rather than confident targets, giving more weight to sources from the original selection function than to the new ones, which is problematic if the starting selection function is too conservative. To address this issue, the selection process after a bootstrap step could be modified to include criteria for adding a source to the confident target list. For example, if a source is confidently detected over multiple consecutive bootstrap steps, it can be considered easy enough to detect. Also, in our current approach, we suppose that a source from the initial selection function is necessarily detectable, which might not be true. We already stated that sources originally selected that do not match the previous model's predictions could be flagged as difficult, and if they are never detected across multiple bootstrap steps, they should likely be removed. This would be particularly useful for applications involving artifacts that can be selected by an initial function based on brightness but that do not match the morphology of typical sources from the training dataset.

\begin{table*}[h]
\centering
\caption{\label{table:poor_select_scores} SDC2 scores for models trained from a degraded selection function.}
\begin{tabular}{ l c c c c c c}
 \hline
 \hline
 Model & $M_s$ (Score) & $N_{\rm det}$ & $N_{\textrm{match}}$ & $N_{\rm false}$ & Purity & $\bar{s}$ \\
 \hline
 B-BASE        & 23\,238 & 34\,688 & 32\,031 & 2664 & 92.34\% & 0.8084 \\
 B-BT1         & 24\,657 & 36\,548 & 33\,612 & 2941 & 91.97\% & 0.8209 \\
 B-BT2         & 24\,978 & 36\,938 & 33\,816 & 3130 & 91.55\% & 0.8310 \\
 \hline
\end{tabular}
\end{table*}

\FloatBarrier

\section{Ancillary models for transfer learning}
\label{sec:alt_models}

In this paper, we followed the original challenge conditions as closely as possible to produce results that can be compared with other teams. Here, we aim to take a step toward generalizing our method to observed hyperspectral datasets by distributing models pre-trained on the MAIN cube. For this purpose, we modified our training pipeline to generate examples from the MAIN cube at each iteration, resulting in a much larger training sample than with the LDEV cube. This change would technically allow one to constrain a more complex backbone and to relax some hyperparameters. However, it would require a tedious new exploration of the associated parameter space. In addition, using the small LDEV catalog as a validation dataset makes it difficult to accurately compare the performance of different setups. For these reasons, we chose to keep the exact setup from our LDEV cube training, which had already demonstrated high detection performance. We used the same starting selection function as for our BASE training and performed two bootstrap steps, producing three MAIN cube models: MC-BASE, MC-BT1, and MC-BT2. 

The inference pipeline works the same way as before, and periodic predictions over the LDEV cube were used to monitor for blatant overtraining. The objectness selection threshold was optimized against the LDEV score for each model. We summarize LDEV scorer-based results in Table~\ref{table:main_cube_train_scores}. Given the small number of LDEV sources, the scores we obtain are difficult to compare and likely sensitive to retraining variability. Still, we observe a clear 5\% score improvement between MC-BASE and MC-BT1, while the increase from MC-BT1 to MC-BT2 is within typical retraining variability. We distribute these three models publicly, along with all other models presented in the paper, at \href{https://doi.org/10.5281/zenodo.18403011}{10.5281/zenodo.18403011}. These MAIN-cube models should be used preferentially when transferring SDC2 pre-trained models to new datasets, as they have been exposed to greater example diversity. They are also the models used in the interactive SDC2 inference notebook in the CIANNA GitHub repository, since inference on the small LDEV cube is faster and easier to fit within limited computing resources.

\begin{table*}[h]
\centering
\caption{\label{table:main_cube_train_scores} LDEV SDC2 scores for models trained on the MAIN cube.}
\begin{tabular}{ l c c c c c c}
 \hline
 \hline
 Model & $M_s$ (Score) & $N_{\rm det}$ & $N_{\textrm{match}}$ & $N_{\rm false}$ & Purity & $\bar{s}$ \\
 \hline
 MC-BASE        & 1179 & 1770 & 1614 & 156 & 91.2\% & 0.827 \\
 MC-BT1         & 1246 & 1761 & 1650 & 111 & 93.7\% & 0.823 \\
 MC-BT2         & 1262 & 1761 & 1652 & 109 & 93.8\% & 0.830 \\
 \hline
\end{tabular}
\end{table*}

\end{appendix}

\end{document}